\documentclass{aa}
\usepackage{natbib}
\usepackage{xspace}
\usepackage{amssymb}
\usepackage{amsmath}
\usepackage{graphicx}
\usepackage{comment}

\newcommand{\changes}[1]{#1}
\newcommand{\changestwo}[1]{#1}

\begin{document}
\title{The inner rim structures of protoplanetary discs}
\titlerunning{The inner rim structures of protoplanetary discs}
\authorrunning{Kama, Min, Dominik}
\author{M.~Kama$^{1}$, M.~Min$^{1,2}$ \& C.~Dominik$^{1,3}$}
\institute{$^{1}$Anton Pannekoek Astronomical Institute, University of Amsterdam, P.O. Box 94249, 1090 GE Amsterdam, The Netherlands; E--mail: M.Kama@uva.nl\\
$^{2}$Utrecht Astronomical Institute, University of Utrecht, P.O. Box 80000, 3508 TA Utrecht, The Netherlands\\
$^{3}$Department of Astrophysics, Radboud University Nijmegen, Postbus 9010, 6500 GL Nijmegen, The Netherlands}
\date{DRAFT, \today} 

\abstract{The inner boundary of protoplanetary discs is structured by the dramatic opacity changes at the transition from the dust-containing to a dust-free zone. This paper explores the variety and limits of inner rim structures in passively heated dusty discs. For this study, we implemented detailed sublimation physics in a fast Monte Carlo radiative transfer code. We show that the inner rim in dusty discs is not an infinitely sharp wall but a diffuse region which may be narrow or wide. Furthermore, high surface densities and large silicate grains as well as iron and corundum grains decrease the rim radius, from a 2.2AU radius for small silicates around a $\rm 47L_{\odot}$ Herbig Ae star typically to 0.4AU and as close as 0.2AU. A passive disc with grain growth and a diverse dust composition must thus have a small inner rim radius. Finally, an analytical expression is presented for the rim location as a function of dust, disc and stellar properties.}

\maketitle


\section{Introduction}

	The inner regions of protoplanetary discs are the birthplaces of terrestrial planets. Understanding their nature and diversity is key to our picture of the formation of planetary systems.

	This paper uses Monte Carlo radiative transfer modelling and detailed sublimation physics to explore the inner rim structures of passive dust discs, and their merits in reconciling theory with observation.

	T Tauri and Herbig Ae/Be stars have prominent infrared excesses in their spectral energy distributions (SEDs), due to the presence of protoplanetary discs. The existence of such discs is known observationally through statistics of photometric colours \citep{AdamsLadaShu1987, RydgrenZak1987, Calvetetal1992, LadaAdams1992}, spectral fitting \citep{KH87}, imaging \citep{McCaughreanODell1996, Burrowsetal1996}, interferometry \citep{ManningsSargent1997}, and is inferred from studies of the Solar System and exoplanets. Based on broadband SEDs, discs are classified as either self-shadowed or flaring \citep{KH87, Meeusetal2001, DullemondDominik2004} and their temperature structure can be approximated with two layers to explain solid-state silicate emission bands \citep{CG97}. In the inner regions of discs, dust sublimates. The location where starlight reaches an optical depth of unity in the dust is referred to as the inner rim. The dust sublimation temperature is around 1500K, so the inner rim radiates in the near-infrared.

	In this paper, we use a new Monte Carlo radiative transfer code to study the effects of various physical parameters on the inner rim structure of protoplanetary discs. Our motivation is outlined in Section~\ref{sec:motivation}. Dust sublimation, backwarming and other theoretical aspects are introduced in Section~\ref{sec:methods} and the results are presented in Section~\ref{sec:results}. A discussion and conclusions follow in Sections~\ref{sec:discussion}~and~\ref{sec:conclusions}.

\section{Observational and physical motivation}\label{sec:motivation}

	Herbig Ae/Be stars typically show excess near-infrared (NIR, $\rm\lambda\approx 1.25\ldots 7\mu m$) emission, corresponding to temperatures of $\rm\sim 1500K$. Most fractional NIR excesses, $\rm f_{NIR}=L_{NIR}/L_{\star}$, form a rising histogram from $\rm f_{NIR}\approx 0.02$ to 0.22 and are well reproduced by models including dust sublimation in a hydrostatic disc \citep[hereafter DDN01 and IN05]{DDN2001, IN05}. Larger excesses, e.g. $\rm f_{NIR}=0.23$ and 0.25 for AB Aur and WW Vul \citep{Nattaetal2001}, and $\rm f_{NIR}=0.32$ for HD 142527 \citep{Dominiketal2003}, have proven impossible to reproduce with existing models without introducing unrealistic assumptions about the rim scaleheight or dust sublimation temperature.

	Herbig Ae/Be and T Tauri stars have been shown to follow a $\rm R_{rim}\propto L_{\star}^{1/2}$ law for inner rim radii measured in the NIR regime \citep{Monnieretal2005, MillanGabetetal2007}, thought to coincide with the dust rim. Most Herbig Ae/Be rims in this sample have been shown to be scattered within expected limits \citep{Monnieretal2005} for a radius determined by dust sublimation, consistent with an optically thin inner hole. Some Herbig Be stars, with $\rm L_{\star} > 10^{3}L_{\odot}$, conform to models where the inner hole is optically thick in the midplane and the rim radii are consequently smaller. Gas undergoing accretion onto the star has been assumed as the inner hole opacity source in these cases.
	
	The presence of a smooth, extended hot emission component interior to the dust sublimation radius has been deduced for several systems from interferometric observations \citep{Eisneretal2007, Tannirkulametal2008a, Tannirkulametal2008b, Krausetal2008, Isellaetal2008}. A hot emission component in the inner hole has recently gained further support from spatially resolved spectroscopy \citep{Najitaetal2009, Eisneretal2009}.

	Meanwhile, the physico-chemical parameter space of dust rim models has been little explored in terms of variations in global inner disc properties (surface density profile, composition) and dust type (astronomical silicate has been the standard opacity and homogeneous spheres the particle shape). The complex geometry and high optical depth of protoplanetary discs can now be studied with fast Monte Carlo radiative transfer codes \citep{Minetal2009}, providing further incentive to incorporate more detailed physics to clarify the limits of dusty rim models.

	\begin{figure}
		\includegraphics[width=1.0\linewidth]{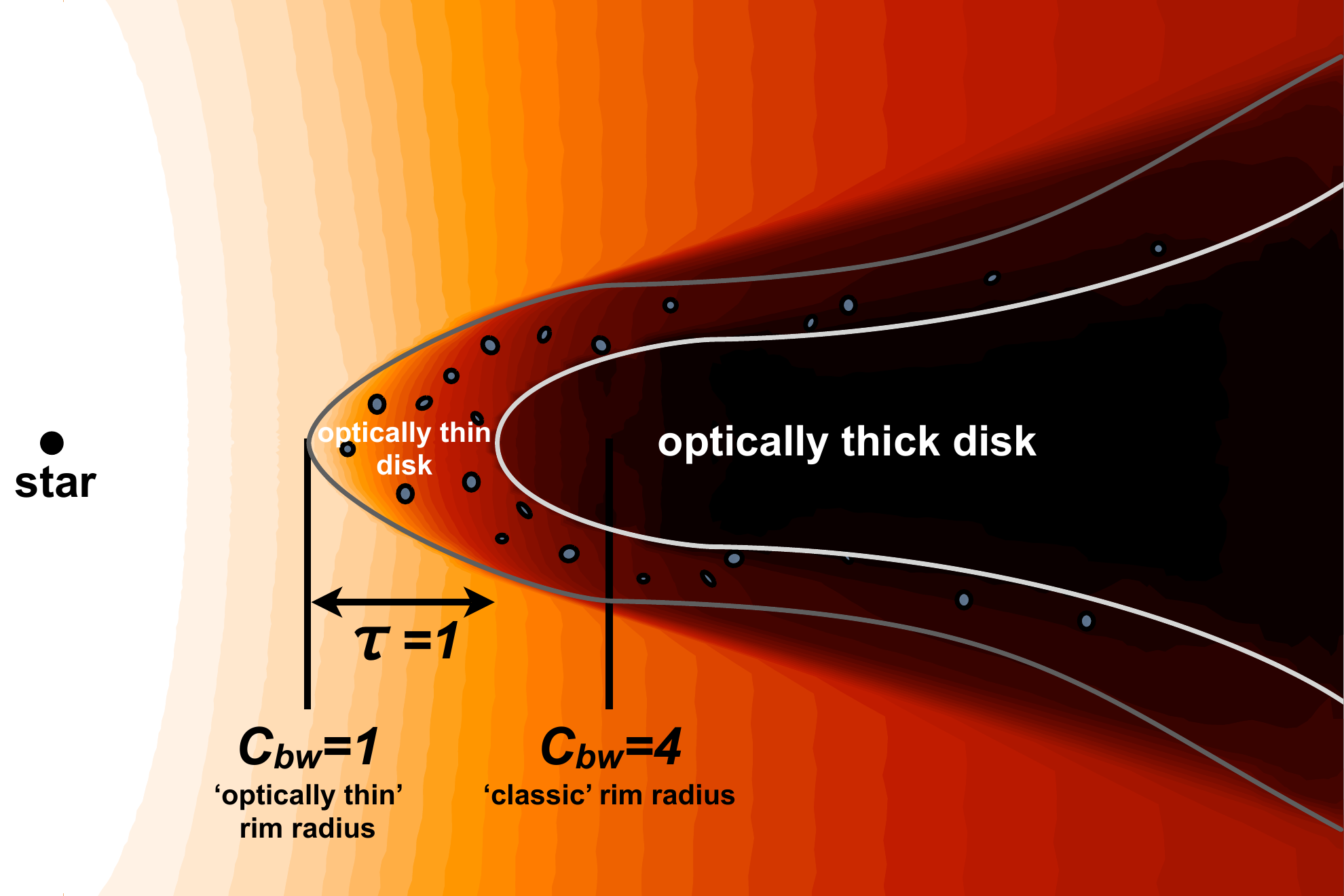}
		\caption{A schematic of the inner few AU of a disc. The colour map indicates the temperature, from $\rm <300K$ (\textit{dark}) to $\rm >2000K$ (\textit{white}). At the optically thin radius, $\rm C_{bw}=1$, dust can begin to condense. The classical rim, at $\rm C_{bw}=4$, is a maximally backwarmed dust wall. The actual rim, i.e. $\rm\tau=1$ surface, is between these limits and preceded by an optically thin zone.}
	\label{fig:rimsketch}
	\end{figure}

\section{Methods}\label{sec:methods}

	We consider models of passive discs, with temperature structures dominated by irradiation from the central star, and no accretional heating. \citep[For a broader review of disc models, we refer the reader to][]{Dullemondetal2007}. The modern paradigm in passive disc models was developed by \citet[hereafter CG97]{CG97}. In this approach, the disc consists of a cool midplane and a hot surface layer. Such models are highly successful in reproducing the general form and solid state bands of young star SEDs, but not the NIR excess, as emission from the dust sublimation region is not considered.

	Any realistic dust species will sublimate at or below a temperature of $\rm T_{subl} \sim 1500K$, corresponding to an emission maximum around $\rm 2\mu m$. In a disc, this implies a radius inside which no dust can exist. Terminating the CG97 density distribution at such a radius results in a dust wall that is exposed to direct starlight, becomes hot and ``puffs up'' \citep[and DDN01]{Nattaetal2001}. Backwarming, explained later, pushes the sublimation location to larger radii. The DDN01 models explain the NIR excess, but the emission is very inclination dependent, an effect contradicted by observations.

	A semi-analytical treatment including a gas density dependent sublimation temperature was applied successfully in explaining observed SEDs and deriving important properties such as characteristic grain sizes \citep[IN05 and][]{Isellaetal2006}. The rim surface was shown to be curved rather than wall-like, due to the dependence of the sublimation temperature on gas density (IN05). \citet{Isellaetal2006} successfully fitted the SEDs and interferometric rim radii of four out of five Herbig Ae stars with IN05 models. In most cases, the best fits required astronomical silicate grains of $\rm \geq 1.2\mu m$ size. However, the modelled rim radii were larger than those obtained by \citet{Eisneretal2004} by up to a factor of three, and simultaneously fitting the SED and rim radius of AB Aur was unsuccessful.

	In a passive disc, the structure of the inner rim is determined by the stellar parameters, the surface density, cooling efficiencies and sublimation temperatures of the dust species, and by backwarming. Because the dust temperature and hydrostatic structure are also intimately coupled in a non-trivial geometry, finding static, stable solutions in terms of dust temperature structure is a difficult task, for which Monte Carlo radiative transfer is well suited.

	Monte Carlo radiative transfer has been used to extend the IN05 models to two grain sizes in thermal contact \citep{Tannirkulametal2007}. It was shown that grain growth, specifically the presence of grains of different sizes and settling of large grains, strongly curves the inner rim. A comparison with these results is made in Section~\ref{sec:discussion}.

\subsection{The bulk gas and vapour densities}\label{sec:densities}

	\changes{The total amount of gas in a disc is referred to as the bulk gas in this paper. The \textit{bulk gas density} determines the amount of dust present if a fixed gas to dust mass ratio, e.g. $\rm f_{gd}=100$, is assumed. Dust grains always contribute some particles to the gas phase. These do not appreciably change the gas mass but do maintain the gas-grain equilibrium, with gas phase particles of the dust species sticking onto a grain and balancing its evaporation. If all of a given dust species must be in the gas phase to maintain this equilibrium, we call its density the \textit{vapour density}. In reality, many gas-phase species may combine to maintain the gas-grain equilibrium of any solid species. As at most a few species exist simultaneously in our models, we generally neglect this fact, with some exceptions (see Sec.~\ref{sec:whereistherim}).}

\subsection{Dust sublimation}\label{sec:sublimation}

	The temperature of a dust grain is determined by its wavelength-dependent opacity, and by the intensity and spectrum of the radiation field responsible for the heating. The cooling efficiency, given by $\rm \epsilon~=~\kappa_{P}^{\prime}(T_{dust})/\kappa_{P}^{\prime}(T_{\star})~=~C_{abs}(T_{dust})/C_{abs}(T_{\star})$, is a ratio between the Planck mean opacity of the dust species at its own temperature to that at the stellar temperature, or for a single grain the ratio of absorption cross sections. In an identical stellar radiation field, a grain with a large $\epsilon$ will achieve radiative equilibrium at a lower temperature than a grain with a small $\epsilon$, and generally epsilon increases with grain size.

	\begin{figure}
		\includegraphics[bb=2.5cm 1.75cm 19cm 25cm, angle=90, clip=, width=1.0\linewidth]{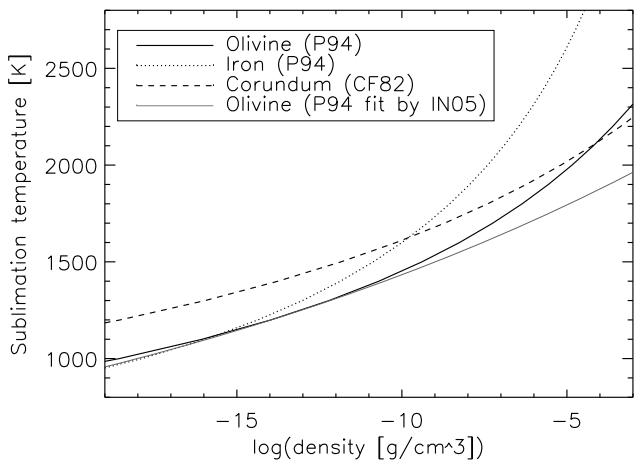}
		\caption{\changes{Sublimation curves of the species used in this work. The vapour density of the species is used. At low vapour densities, corundum is the most refractory, whereas at higher values iron is relatively more refractory. At least one of the two species is always relatively more refractory than olivine. The expression for the curves is given in Eq.~\ref{eq:CCevaplaw}, and the fitting parameters in Table~\ref{tab:TfitABnew}. The IN05 fitting formula has been abundance-corrected for this plot, otherwise it would lie $\rm\sim 50K$ below our curve for any olivine vapour density (see Appendix~\ref{apx:differences}). For astronomical silicate and forsterite, the olivine sublimation curve was used in modelling.}}
	\label{fig:sublimation}
	\end{figure}
	
	For a grain to be stable, the particle flux excaping from its surface must be balanced by the flux coming onto it from the gas phase. If the temperature gets too high, the grain will sublimate. The first rim models assumed that this would occur at $\rm T_{subl}=\mathrm{const}$, e.g. DDN01 used 1500K. Generally, the sublimation temperature of a dust species is a function of the partial pressure of its gas phase component, $\rm T_{subl} = \mathrm{f}(P)$. \changes{The ideal gas equation of state can be used to replace the partial pressure P of the species with a function of its temperature and its vapour density, $\rm\rho_{vapour}$.} Loosely following \citet{Zemansky1968}, the vapour density at saturation pressure is given by the Clausius-Clapeyron equation:

	\begin{equation}
		\rm \log{(\rho_{vapour})} = B - \frac{A}{T_{subl}} - \log{\left(T_{subl}^{-C}\right)}\textrm{,}
	\label{eq:CCevaplaw}
	\end{equation}

	where A, B and C are thermodynamical quantities that can be fitted using laboratory measurements\footnote{
	Thermodynamically, $\rm A = \Delta H /\ln{(10)}$, where $\Delta H$ is the enthalpy of vapourization, and $\rm C = (c_{0}^{\prime\prime\prime}/R -1)$. The molar heat capacity has been broken into constant and temperature-dependent components, $c_{P}^{\prime\prime\prime} = c_{0}^{\prime\prime\prime} + c_{i}^{\prime\prime\prime}$. In addition, $\rm B = B_{1} + \log{(k/\mu m_{p})}$, where B$_{1}$ is given by

	\[
		\rm B_{1} = \frac{1}{\ln{(10)}} \left( \int_{0}^{T}\left[ \frac{\int_{0}^{T}c_{i}^{\prime\prime\prime}dT - \int_{0}^{T}c_{P}^{\prime}dT}{T^{2}R} \right]dT + {I} \right) \textrm{,}
	\]

	and I is an integration constant. Sometimes, for example in \cite{Smithells1967} as referenced by \cite{Lamy1974}, the full form of Eq.~\ref{eq:CCevaplaw} must be invoked to provide a sufficiently accurate fit to experimental data, however most often, as in this paper, the assumptions $B \neq B(T)$ and $C=-1$ are made.
	}.

	We take $\rm C=-1$ and obtain A and B from a published table of evaporation temperatures with corresponding bulk gas densities and abundances \citep[][P94]{P94}. The resulting parameters for Eq.~\ref{eq:CCevaplaw}, alongside those obtained from other sources, are summarized in Table~\ref{tab:TfitABnew}. In fitting the P94 data, the partial pressure of each species was expressed through its abundance and molecular mass, and the bulk gas density.  Parameters from other sources were converted (see Eq.2 of \citet{Lamy1974} and Eq.10 of \citet{CameronFegley1982}). The table also gives $\rm T_{subl}$ at $\rm\rho_{vapour}= 10^{-12} g/cm^{3}$ and references for each species, with boldface denoting the species used in this study.
	
	Sublimation curves of the main species used in this work are shown in Fig.~\ref{fig:sublimation}. At vapour densities of $\rm 10^{-12\ldots-10} g/cm^{3}$, olivine sublimates at around 1300K, iron at 1400K and corundum at 1500K. Corundum must have an abundance $\rm<1\%$ of the total dust mass and thus will necessarily have a much lower partial density than olivine or iron.

\begin{table*}[!ht]
\begin{center}
\begin{tabular}{|r|l|r|r@{.}l|r@{.}l|l|l|}
\hline
Species & Chemical & $\rm T_{subl}$ & \multicolumn{2}{c|}{A} &
\multicolumn{2}{c|}{B} & Reference for A and B & Reference for opacity\\
& formula & [K] &  \multicolumn{2}{c|}{(Eq.~\ref{eq:CCevaplaw})} &  \multicolumn{2}{c|}{(Eq.~\ref{eq:CCevaplaw})} & &\\
\hline
Astron. silicate 	& MgFeSiO$_{4}$	& 	1300	& 28030	&	& 12	& 471	& \citet[P94]{P94}		&  \citet{DraineLee1984}\\
Olivine	& MgFeSiO$_{4}$		&	1300	& 28030	&	& 12	& 471	& P94		& \citet{Dorschneretal1995}\\
Forsterite	& Mg$_{2}$SiO$_{4}$	&	1150	& 26091	&	& 13	& 418	& \citet[CF82]{CameronFegley1982}		& \citet{1973Pssb...55..677S}\\
Iron	& Fe				&	1400	& 21542	&	& 6	& 6715	& P94		& P94\\
Iron$^{\star}$		& Fe				&	1150	& 20686	&	& 9	& 1134	& CF82		& \\
Corundum	& Al$_{2}$O$_{3}$	&	1500	& 40720	&	& 18	& 479	& CF82		& \citet{1995Icar..114..203K}\\
Orthopyroxene$^{\star}$		& MgSiO$_{3}$		&	\ldots	& 30478	&	& 14	& 898	& P94		& \\
Quartz$^{\star}$		& SiO$_{2}$			&	\ldots	& 26335	&	& 11	& 184	&  \citet{Schick1960, Lamy1974}	& \\
Water ice$^{\star}$			& H$_{2}$O	&	150	& 2827	& 7	& 7	& 7205	& P94		& \\
Troilite$^{\star}$		& FeS			&	680	& 155	& 91	& -4	& 9516	& P94 (note: $\rm T_{subl}=const$)		& \\
\hline
\end{tabular}

\label{tab:TfitABnew}
\end{center}
\caption{A summary of the dust properties. ($^{\star}$  -- not used in the present work, given for completeness.)}
\end{table*}

\subsection{Backwarming}\label{sec:backwarming}

	\begin{figure}
		\includegraphics[clip=, width=1.0\linewidth]{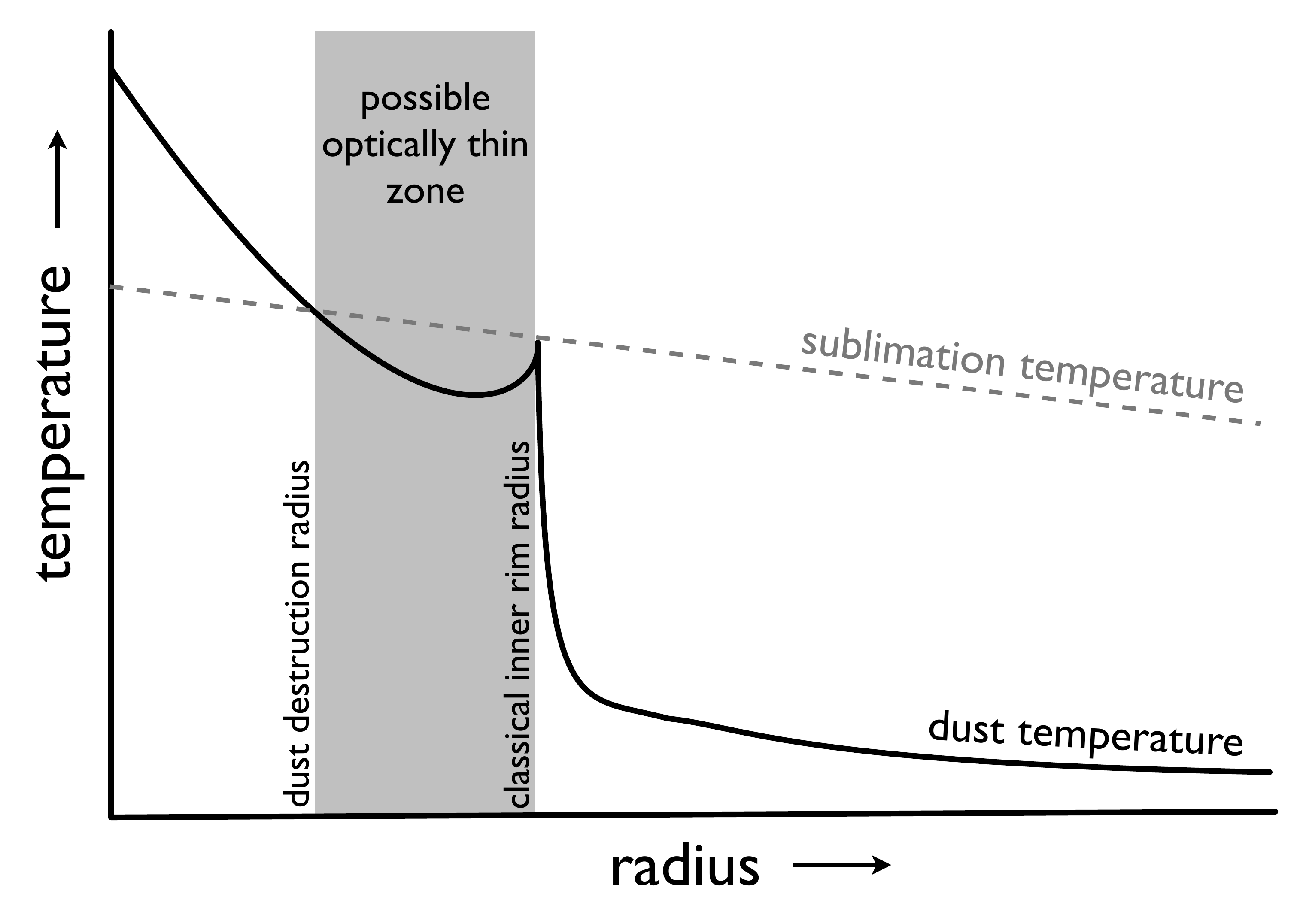}
		\caption{A schematic midplane dust temperature profile of an efficiently backwarmed rim. In front of the rim, the optically thin equilibrium temperature of the dust grains is plotted. Once this drops below the sublimation temperature, dust can condense. \changes{Generally, the sublimation temperature decreases with radius because of the increasing scaleheight and decreasing surface density.} Between the dust destruction and classical rim radii, there is a region in which dust may condense but need not be optically thick. Dust in this region is warm, but not close to the sublimation temperature. The dust temperature peaks locally at the optically thick rim location, and then falls rapidly to a low value inside the disc. See also Fig.~\ref{fig:rimsketch}.}
	\label{fig:Tprofile}
	\end{figure}

	If a dust grain cannot cool into empty space in every direction, its temperature increases due to backwarming. Geometrically, the backwarming factor $\rm C_{bw}$ can be defined as the ratio of a full $\rm 4\pi$ solid angle to the solid angle subtended by the empty sky seen by the grain. For each photon emitted into the sky area obscured by nearby matter, a similar photon returns, hence the term ``backwarming''. An estimate of the temperature of a dust grain on the rim surface is obtained from the equation

	\begin{equation} 
		\rm \pi \frac{L_{\star}}{4\pi R_{rim}^{2}} = \frac{4\pi}{C_{bw}}\epsilon \sigma T_{d}^{4},
	\label{eq:BW}
	\end{equation}

	where the energy absorbed by a grain at a distance $\rm R_{rim}$ from a star of luminosity $\rm L_{\star}$ is equated with that emitted by it, assuming a cooling efficiency $\epsilon$. \changes{Rigorously, $\rm C_{bw}$ is wavelength-dependent and its value can be obtained by solving a complex radiative transfer and dust sublimation and condensation problem, which is done in the Monte Carlo code we employ. However, as we show later, the description in Eq.~\ref{eq:BW}, while simple, is useful in interpreting the results of our more rigorous models,} as well as other work on the subject. Other things being constant, the rim location scales as $\rm R_{rim}\propto C_{bw}^{1/2}$.

	$\rm C_{bw}=1$ yields the temperature of a grain in an optically thin environment, and when $\rm T_{subl}$ is considered it gives the smallest possible distance from the star a passive grain may orbit \changestwo{for longer than its sublimation timescale}. \changes{This limit has been used by e.g. Tuthill (2001) and Monnier \& Millan-Gabet (2002) to obtain lower limits \changestwo{on disc inner radii. It must be kept in mind that such limits are derived under assumptions regarding grain opacities, sublimation temperatures and geometry.}} In assuming the rim was an instantaneously optically thick vertical wall, DDN2001 adopted $\rm C_{bw}=4$. This illustrates how far backwarming can move the rim. As Monnier et al. (2005) pointed out, the rim is actually between the two limits, depending on over what distance the dust becomes optically thick. The concept of such an optically thin region in front of the rim is illustrated by Figures~\ref{fig:rimsketch}~and~\ref{fig:Tprofile}.

	On the rim, $\rm C_{bw} > 4$ may also be true for special locations. One can imagine this for a dust grain sitting at the far end of a tunnel which extends into the rim (see also Appendix~\ref{app:numerics}).

\subsection{Monte Carlo radiative transfer with MCMax}

	Radiative transfer through the complex dust geometries we wish to consider requires a flexible method in terms of density structures and dust composition. In addition, the coupled nature of the dust temperature and the hydrostatic structure also calls for an iterative solution. The required flexibility and speed can be obtained by using Monte Carlo radiative transfer.

	We use here the Monte Carlo radiative transfer code MCMax \citep{Minetal2009}. Because of an efficient implementation of the diffusion approximation for regions with high optical depth, this code is especially suited for fast computations of geometries involving high optical depths.

	The radiative transfer is based on the Monte Carlo scheme of continuous reemission developed by \citet{BW2001} and optically thin regions are treated with the method of \citet{Lucy1999}.

	To model the inner rim, dust sublimation was implemented into MCMax according to Eq.~\ref{eq:CCevaplaw}. The density structure of the dust is solved for by iterating the radiative transfer, dust sublimation and recondensation, and vertical hydrostatic equilibrium. After each iteration, the spatial grid is rebuilt in order to properly sample the new density structure. This ensures that in each iteration the radial resolution around $\rm\tau\approx 1$ is sufficient for $\rm\Delta\tau$ through any cell to be less than a fraction of unity.

	Condensation, if it can occur, must be treated carefully. An upper limit is to transfer all the gas phase dust into grains, but if done too quickly this may overheat the cell, forcing sublimation and thus a loop. In the rim region, between $\rm C_{bw}=1$ and 4, the change in solid dust mass \changes{in any cell} is weighted so that \changes{the change is} smaller the closer the cell is to the local $\rm T_{subl}$. The gas fraction of any dust species is constrained to be radially and vertically monotonous in this region. Furthermore, the optical depth added radially in front of the $\rm\tau=1$ surface cannot exceed 0.1. Some physical solutions might be suppressed by our constraints, but we expect a proper treatment of these to require time-dependent modelling. For a detailed discussion of the numerical aspects of implementing the sublimation physics we refer the reader to Appendix~\ref{app:numerics}.

	In reality, the sublimation of grains will cause the grains to lose mass, and thereby decrease in size. For practical considerations, and because we want to systematically study the effects of varying the grain size, we keep the size of the grains independent of the sublimation state, but increase or decrease their number density.

\subsection{Disc parameters}

	\subsubsection{General properties}

	The disc model is parametrized as follows: a radial dust surface density distribution $\rm\Sigma(R)$ is described by two joined power laws, $\rm R^{-p_{1}}$ and $\rm R^{-p_{2}}$, extending from an inner radius $\rm R_{in}$ out to $\rm R_{out}$. The power laws are joined at the location $\rm R_{p}$. The total dust mass, $\rm M_{dust}$, gas to dust ratio, and a dust composition are specified.

	In general, we assume $\rm R_{in}=0.03AU$, $\rm R_{out}=200AU$, $\rm p_{1}=0.0$, $\rm p_{2}=1.5$ and $\rm R_{p}=4AU$. A flat inner disc surface density makes the interpretation of our results more straightforward, especially for the optically thin zones.

	Most of our models assume a dust mass of $\rm M_{dust}\approx 10^{-4}M_{\odot}$. For a parameter study of the inner disc surface density, this is varied from $\rm 10^{-7}M_{\odot}$ to $\rm 10^{-2}M_{\odot}$ ($\rm 10^{-1}M_{\oplus}$ to $\rm 10^{4}M_{\oplus}$).

	\changes{The gas to dust ratio is 100 everywhere, and the gas and dust scaleheights are coupled.} All our models use the star previously employed by IN05, with $\rm T_{\star} = 10000K$, $\rm L_{\star} = 47L_{\odot}$, and $\rm M_{\star}=2.5M_{\odot}$.

	The IN05 comparison discs are parametrized as $\rm R_{in}=0.1AU$, $\rm R_{out}=200AU$ and $\rm p1=p2=1.5$, with homogeneous spheres (HS) of astronomical silicate as the opacity model.

	\subsubsection{Dust types}

	The opacities of the dust grains are sensitive to their size and shape. For modelling the effects of the grain shape on the opacities, we use the statistical approach \citep[][]{BohrenHuffman, Minetal2003}. This allows to take the grain shape effects into account properly using computationally favorable grain geometries. We apply the distribution of hollow spheres \citep[DHS; see][]{Minetal2005}. 

	For the composition of the grains previous studies have frequently used the so-called astronomical silicate \citep{DraineLee1984}. This species is most likely a composite of different materials with different sublimation laws. Its composition is unknown and we are thus unable to derive a proper sublimation law for astronomical silicate. Therefore, in this paper we focus mainly on well characterized materials which have been measured in the laboratory and are known to exist in meteorites.

	We begin by exploring discs with various grain sizes and surface densities of amorphous olivine. Then, models with two olivine grain sizes, and with dust types of different transparency, refractivity and cooling effiency are presented.

	The presence of small, $\rm\sim 0.1\mu m$ grains in discs is revealed by fitting observed solid state emission bands. We use a model with such grains as a reference point. Large, $\rm\geq 10\mu m$ grains have a higher cooling efficiency than small grains, and as grain growth is known to occur in discs, they are also of great interest in modelling. To study the effects of large grains on the inner disc structure, we present models with 10 or $\rm 100\mu m$ olivine grains added to $\rm 0.1\mu m$ particles. Discs of 1, 10, 100 and $\rm 1000\mu m$ olivine are presented in Sections~\ref{sec:results}~and~\ref{sec:discussion}.

	Surface density influences the inner rim in several ways: it sets the sublimation temperature, influences the backwarming efficiency and determines over what radial distance any given optical depth is reached. We vary the dust surface density in the inner disc from $\rm\Sigma_{d}=10^{-4}$ to $\rm 10^{3}$ as a parameter study.

	In modelling two grain types, we assume they are not in thermal contact, unless the contrary is explicitly stated. This is to allow more efficiently cooling or refractory species to exist independently of less efficiently cooling or more volatile species. Thermal coupling will arise from the Monte Carlo radiative transfer in sufficiently dense regions, where all species will be in radiative thermal equilibrium. Unless stated otherwise, abundance ratios of 9999/1, 90/10 and 10/90 are used in studying grain mixtures.

	Above $\rm\sim 1000K$, amorphous silicates anneal to form crystalline silicates. The amorphous component is represented by olivine in our models. It has been found that the annealed silicates are often iron-poor and magnesium-rich, e.g. forsterite and enstatite. Observations \citep{vanBoekeletal2004} and the short timescale of crystallization \citep{Lenzunietal1995} suggest that the crystalline silicate fraction on the inner rim surface approaches unity.

	We only present $\rm 10\mu m$ forsterite grains in this paper. Given the expected high abundance of crystalline particles in the inner disc, we also perform full scattering computations for our olivine and forsterite mixtures. Using a $\rm 0.1\%$, $\rm 10\%$ or $\rm 90\%$ forsterite abundance allows to follow the transition from an amorphous- to crystalline-dominated rim, assuming the former is represented by olivine and the latter by forsterite.

	Iron grains, plausibly a sink of iron left over from forsterite formation, is another component we explore. Small iron grains cool more efficiently than small olivine grains, furthermore iron is significantly more refractory in the high vapour density regime, as seen in Fig.~\ref{fig:sublimation}. Abundance ratios of 9999/1 and 90/10 are used in our olivine and iron disc models.

	Corundum has been suggested to exist in discs \citep[e.g.][]{Lenzunietal1995}. It is very refractory in the low vapour density regime, and forms a ``highly refractory envelope'' around olivine together with iron in $\rm T_{subl}$-$\rm\rho_{vapour}$ parameter space, as seen in Fig.~\ref{fig:sublimation}. We use a trace olivine/corundum mass abundance of 9999/1 and compare it with a model where this ratio is 99/1. \changes{The latter is a convenient limit, approaching one obtained by taking a mixture with cosmic abundances and putting all aluminium into corundum.}

	Materials such as olivine do not exist in single molecule form and decompose during sublimation \citep{Duschletal1996}. Thus, particles transferred to the gas phase from one grain species may contribute to upholding the gas-grain equilibrium of another species \citep{Dominiketal1993}. We have attempted to take this into account by using the summed partial pressures of amorphous olivine and crystalline forsterite to uphold the gas-grain equilibrium of both. Any pure iron released by the sublimation of olivine is not considered as a separate dust species in these models.

	For comparison with previous studies we also present several models using astronomical silicate, in these cases the olivine sublimation law is applied.

\section{Results}\label{sec:results}

	\changes{The results of our disc modelling are presented in this section. We first put forward single grain type models with realistic opacities. Then, dependencies on grain size and surface density are discussed. Finally, the effects of adding various amounts of different grain sizes or types are shown. A detailed comparison with the results of IN05 and \citet{Tannirkulametal2007} may be found in Appendix~\ref{apx:differences}.}

	\changes{The optical depths are computed using the dust Planck mean opacity at the stellar temperature of 10000K.}

	\subsection{Olivine grains of various sizes}

	\changes{For the main batch of models, we used opacities of iron-containing olivines, with sublimation properties from P94.}

	\subsubsection{The developing rim}

\begin{figure}
		\includegraphics[bb=2.5cm 1.75cm 19cm 25cm, angle=90, clip=, width=0.95\linewidth]{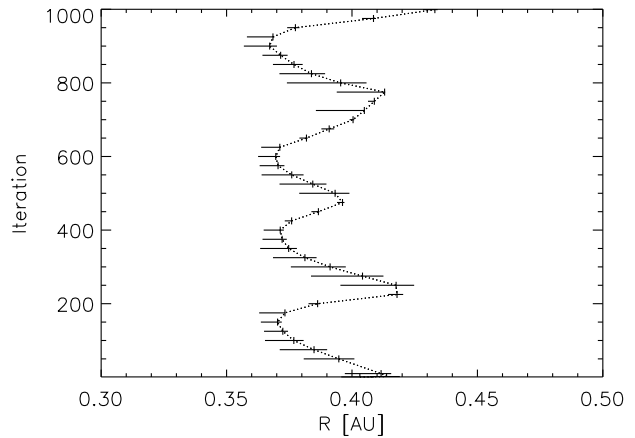}
		\caption{\changes{Distances along the midplane, in AU, where $\rm\tau=0.1, 1$ and 10 is reached, versus iteration number. The evolution of an $\rm\Sigma_{dust}=25 g/cm^{2}$ inner disc with $\rm 10\mu m$ olivine over $10^{3}$ iterations reveals the oscillatory long-term behaviour of our models. Dust is built up over $\rm\sim 150$ iterations, after which the sublimation temperature is exceeded and a retreat occurs, during which the optically thin zone is narrow. The fractional extent of the rim radius variation is significantly smaller for smaller grains.}}
	\label{fig:iterations}
	\end{figure}

	The first important result is that the inclusion of detailed condensation and sublimation physics leads to a rim location which is not entirely stable. This is demonstrated in Fig.~\ref{fig:iterations}, where we show the location and sharpness of the rim as a function of the computational iteration number.

	We begin the computation by placing the rim at the $\rm C_{bw}=1$ location (iteration 0), followed by two backwarming iterations where the location is dictated by an upper limit on $\rm C_{bw}$. The radiative transfer (temperature determination across the disc), vertical structure of the disc, and dust sublimation and condensation are iterated on 50 \changes{to} 200 times. For each iteration, we plot a horizontal bar in Fig.~\ref{fig:iterations}, showing the radii where the optical depth radially along the midplane ranges from 0.1 to 10, with a mark on the $\rm\tau=1$ location. \changes{Thus a short, point-like extent of the bar indicates a sharp rim that jumps from low to high optical depth very fast, i.e. the classical structure as proposed by DDN01.}

	\changes{Over 1000 iterations, we see that both the rim location as well as its inward motion are unstable. Around iteration 150, the sublimation temperature is exceeded in a small region in the rim, which becomes very sharp, with the $\rm\tau=0.1$ and 10 locations moving close together. The backwarming efficiency increases and the rim recedes. Around iteration 250, a new cycle begins with optically thin zone formation.}

	Similar variability in the rim location is evident in models with multiple grain types. The variability in rim structure which we have demonstrated can, in some cases, considerably affect the fractional NIR excess, $\rm f_{NIR}$, making it difficult to associate illustrative values with a particular model. We are working on clarifying the issue.

	It must be emphasized that while we have just described the changes in rim structure as if time-dependent, our modelling has no physical time dependence. Rather, our results seem to indicate that there is no unique, \changes{static} solution for the dust distribution in the rim, at least when a single grain type is used, and variations should be expected in a proper time-dependent treatment.

	\subsubsection{Grain size and surface density}

\begin{figure*}[!ht]
		\includegraphics[bb=0.5cm 0.5cm 18cm 13cm, clip=, width=1.0\linewidth]{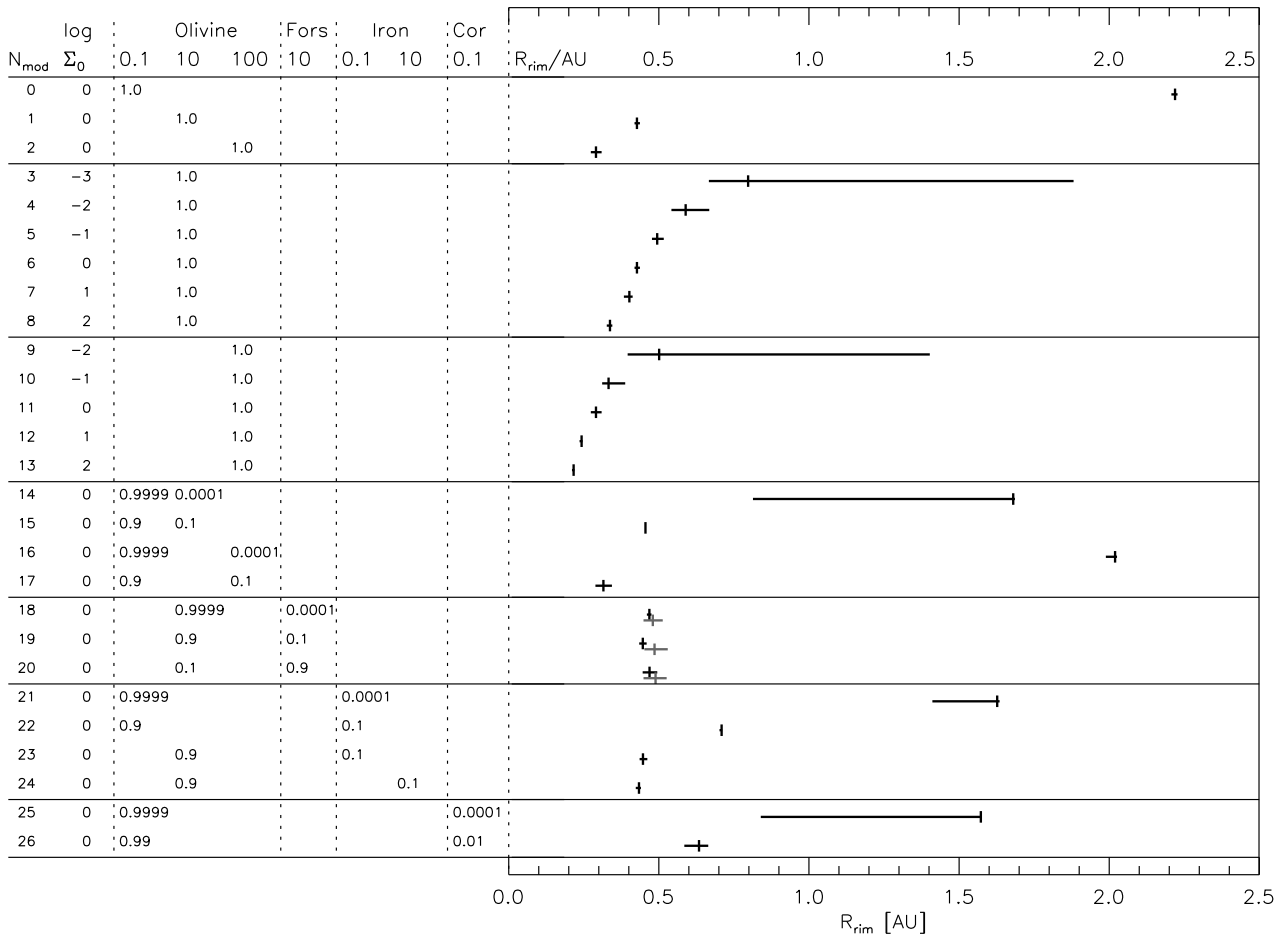}
		\caption{$\rm\tau=0.1$, 1 and 10 radii of selected models. Total dust surface densities are given, and normalized abundances are tabulated. Cases with no scattering are given (\textit{black, upper lines}), with isotropic scattering indicated for forsterite (18-20, \textit{dark gray, lower lines}). The horizontal bars indicate the $\rm\tau=0.1$ and 10 radii. Models $\rm 0\ldots 2$ provide a comparison to surface density ($\rm 3\ldots 13$), size ($\rm 14\ldots 17$) and composition ($\rm 18\ldots 26$) variations.}
	\label{fig:twospecies}
	\end{figure*}

	As we saw in Section~\ref{sec:methods}, both the grain size as well as the surface density influence the rim location. This is demonstrated in the first three panels of the summary diagram in Fig.~\ref{fig:twospecies}, which show the rim location and sharpness for models with three different grain sizes and a range of surface densities. \changes{For a given total (solid and vapour) dust surface density of $\rm 1 g/cm^{2}$ (models 0-2), the rim of $\rm 0.1\mu m$ grains is at 2.22AU, while $\rm 10\mu m$ grains put the rim at 0.43AU and $\rm 100\mu m$ grains at 0.29AU.} A shift of the rim location towards the star is also evident for surface density variations with a fixed grain size of $\rm 10\mu m$ (models 3-8) and of $\rm 100\mu m$ (models 9-13). We will return to this point in Section~\ref{sec:discussion} and show how this can be understood analytically.

	Another feature of the single grain type models is the relative sharpness of the rims. It seems that if enough condensible material is available, the rim will take a sharp form, as assumed in the classical rim models. The low surface density computation in model 3 reveals the situation in discs with low amounts of condensible material available. \changes{In model 3, the $\rm\tau=0.1$ to 10 region extends from 0.67 to 1.88AU because the opacity is too small for the optical depth to rise quickly with radius even if all condensible material is put in the solid phase.} A similar effect is seen in models 9-13, where the same parameter study as with models 3-8 is carried out with $\rm 100\mu m$ grains.

	While in these models, the low surface density or very large grain size might seem artifical, effects similar to those seen here also emerge in multi-grain models with trace species that might be present in the very hot inner regions of discs. We will now turn our attention to these models.

	\subsection{Two sizes of olivine grains}

	Increasing the grain size generally allows cooling to gain in efficiency compared to heating. We will now present the effects of this in models where small and large grains are mixed.

	\changes{Models 14-17 of Fig.~\ref{fig:twospecies} contain two different grain sizes of the same material, in this case olivine. How the optically thin region is cooled by the large olivine grains, and then filled up as the abundance is increased, is demonstrated sequentially as temperature maps in Fig.~\ref{fig:3maps-oo}.}

	\changes{In the upper row of Fig.~\ref{fig:3maps-oo}, the $\rm 0.1\mu m$ grains (left panel) have a rim at 2.22AU, the $\rm 10\mu m$ particles (right panel) at 0.43AU. Differences in dust temperature throughout the inner disc are evident.}

	\changes{Model 14, seen in the middle row, is a disc with a $\rm 0.1\mu m$ to $\rm 10\mu m$ grain mass ratio of 9999/1. The trace abundance of efficiently cooling large grains is sufficient to give rise to an extended, cool region of low optical depth, which shifts the condensation location of $\rm 0.1\mu m$ grains from 2.22AU (model 0) to 1.68AU. The $\rm\tau=10$ location coincides closely with this point, indicating that a large amount of $\rm 0.1\mu m$ grains exists, pushing up the opacity. An optical depth of $\rm\tau=0.1$ is reached at 0.81AU, indicating the optically thin region covers $\rm 50\%$ of the rim radius and around $\rm 75\%$ of the surface inside that radius.}

	\changes{The effect of pulling in the rim is more pronounced in model 15, seen in the bottom row of Fig.~\ref{fig:3maps-oo}, where $\rm 10\%$ of the dust mass is in $\rm 10\mu m$ olivine grains and the rest in $\rm 0.1\mu m$ grains. The inner rim is now at 0.45AU. The large grains dominate the structure, which is close to that of the disc with only large grains (model 1). The smaller, $\rm 0.1\mu m$ grains exist all the way to the new rim location, in a region where they could never condense without the presence of the larger particles.}

	\changes{Model 16 shows that a fractional mass abundance of $10^{-4}$ of $\rm 100\mu m$ grains has very little effect on the small grain rim, for their low abundance is insufficient to contribute to the optical depth. Model 17 demonstrates that a $\rm 10\%$ abundance of $\rm 100\mu m$ grains is sufficient for them to dominate the rim location at a total dust surface density of $\rm 1g/cm^{2}$.}

	\subsection{Olivine and forsterite}

	Forsterite is expected to be very abundant in the inner rim region. We now present modelled disc structures obtained with no scattering (our usual assumption) and with isotropic scattering for mixtures of $\rm 10\mu m$ olivine and forsterite.

	Since relative to olivine, forsterite grains are highly transparent to stellar radiation, one might imagine them to exist very close to the star. In addition to having a low heating efficiency in the stellar radiation field, however, forsterite grains cool very inefficiently at temperatures of 1300$\ldots$1500K. This causes them to sublimate at approximately the same radius as the iron-rich silicates, leading to no difference between the rim structure of model 1 and models 18 to 20 in the case with no scattering, as seen in Fig.~\ref{fig:twospecies}. Including isotropic scattering (lower, dark gray lines of models 18 to 20) slightly increases the rim radius because of the increased backwarming efficiency. Note that at temperatures of a few hundred kelvin, the forsterite grains will cool much more efficiently and be much cooler than the iron-containing silicate grains.

	\subsection{Small amounts of effienctly cooling or ultra-refractory material}

	Small, $\rm 0.1\mu m$ grains of iron and corundum are more efficiently cooling than olivine of  the same size, and may exist significantly closer to the star. The high refractivity of corundum and iron provides another mechanism that allows grains to survive near the star.

\begin{figure*}
		\includegraphics[bb=1.25cm 1.75cm 19cm 25.5cm, angle=90, clip=, width=1.0\linewidth]{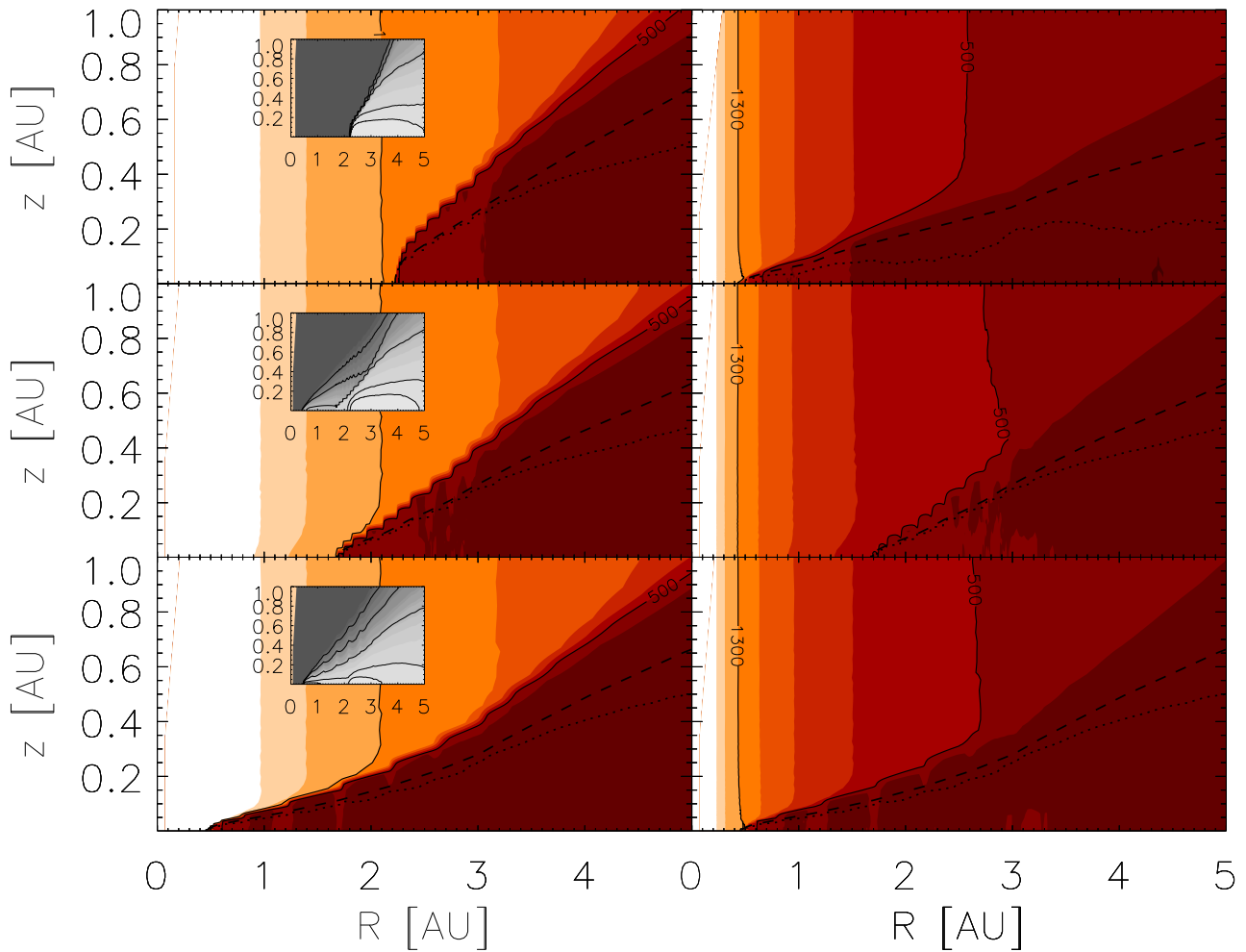}
		\caption{\changes{Temperature (main) and density (insets) maps of disc inner regions for various olivine mixtures. The axes show radial distance from the star along the midplane and vertical height from it. The colour levels represent the dust temperature in steps of 100K, for reference we give solid lines at 500K and 1300K. Also shown are the radial (dashed) and vertical (dotted) $\tau=1$ surfaces for the stellar radiation. The inset maps have total solid dust density contours at factors 2.7, 10, $10^{4}$, $10^{7}$ and $10^{10}$ below the maximum. The dust surface density in these models is $\rm\Sigma_{g}\approx 1 g/cm^{2}$. A typical midplane dust density is $\rm 3\cdot10^{-13} g/cm^{3}$. In models with several dust species, the species are thermally decoupled, but radiative thermal equilibrium arises naturally deep inside the disc. Non-uniformities in the temperature maps reflect Monte Carlo noise and the resolution of the angular grid. \textbf{Top row:} Models of discs with one grain size. Shown is the dust temperature for discs with 0.1 (left panel) and $\rm 10\mu m$ (right panel) olivine grains (models 0 and 1 in Fig.~\ref{fig:twospecies}). \textbf{Middle row:} A disc with a small to large grain mass ratio of 9999/1 (model 14). Given are the temperatures of the small, $\rm 0.1\mu m$ grains (left) and the larger, $\rm 10\mu m$ grains (right). \textbf{Bottom row:} A disc with a mass abundance ratio of 90/10 in small/large grains (model 15). We again give the temperatures of the $\rm 0.1\mu m$ grains (left) and $\rm 10\mu m$ grains (right) that co-exist in this disc.}}
	\label{fig:3maps-oo}
	\end{figure*}

\begin{figure*}[!ht]
		\includegraphics[bb=1.25cm 1.75cm 19cm 25.5cm, angle=90, clip=, width=1.0\linewidth]{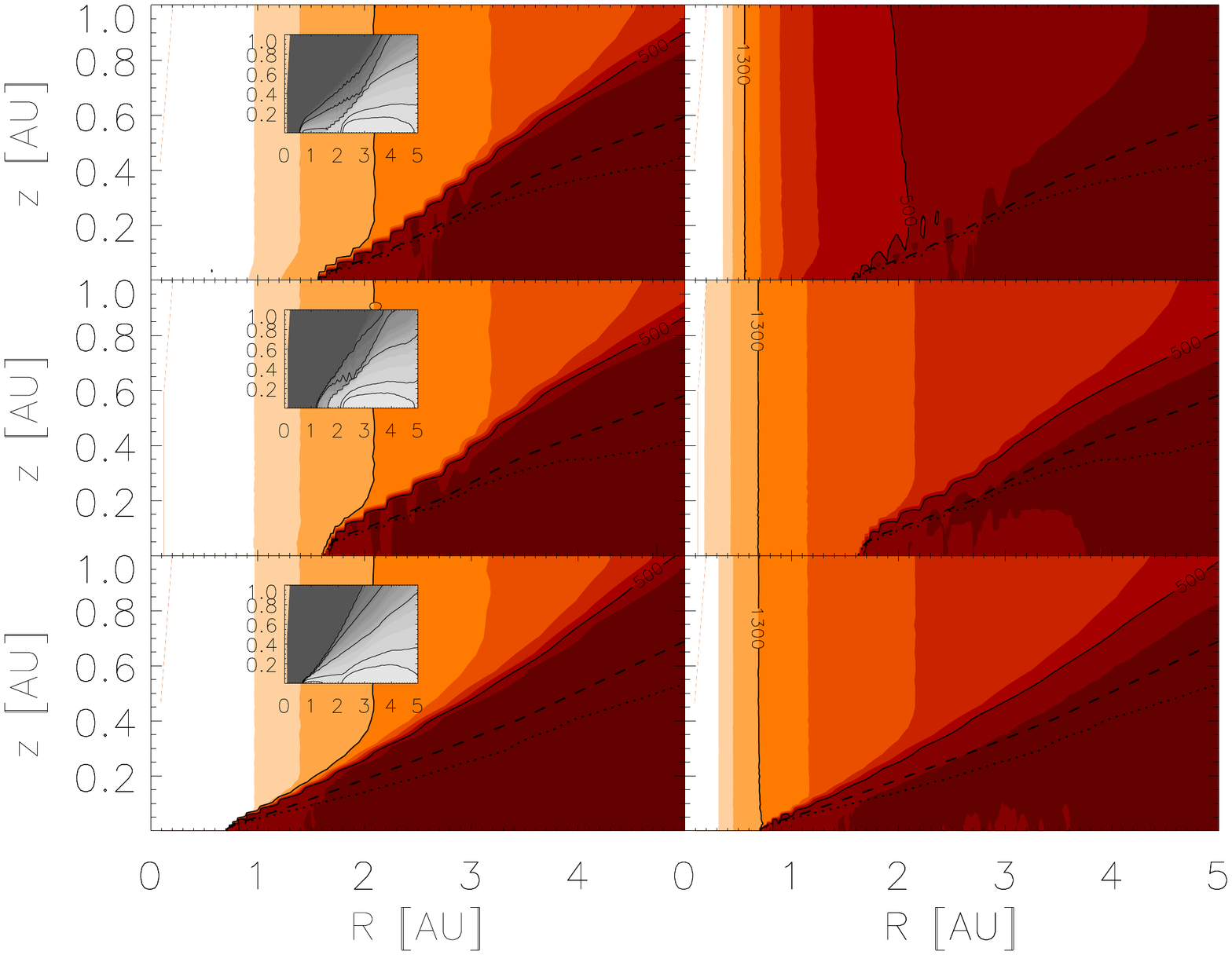}
		\caption{\changes{Temperature (main) and density (insets) maps of disc inner regions for various grain mixtures. See Fig.~\ref{fig:3maps-oo} for a detailed legend. \textbf{Top row:} A disc with an olivine to corundum mass abundance ratio of 9999/1. Shown are the temperatures of the $\rm0.1\mu m$ olivine (left) and the $\rm0.1\mu m$ corundum (right) grains. (Model 25 of Fig.~\ref{fig:twospecies}, compare with model 0.) \textbf{Middle panel:} A disc with an olivine/iron mass abundance ratio of 9999/1 (model 21). Shown are the temperatures of the $\rm0.1\mu m$ olivine (left) and iron (right) grains. \textbf{Lower panel:} A disc with an olivine/iron mass abundance ratio of 90/10 (model 22). Given are the temperatures of the $\rm0.1\mu m$ olivine (left) and iron (right) grains.}}
	\label{fig:3maps-oc}
	\end{figure*}

	\changes{A mass abundance of 0.0001 in $\rm 0.1\mu m$ iron grains pulls the rim of similarly sized olivine from 2.22AU to 1.63AU and gives rise to an optically thin zone covering $\rm\approx 10\%$ of the rim radius (model 21). While $\rm0.1\mu m$ iron is cooler than olivine, it does not cool the inner disc as efficiently as corundum does (see below and Fig.~\ref{fig:3maps-oc}).} A $\rm 10\%$ abundance of small iron pulls the rim to 0.71AU, but now the rim is very sharp (model 22), and the structure resembles that obtained with a 90/10 mass ratio of small and large olivine grains (model 15, see Fig.~\ref{fig:3maps-oo})

	Adding an abundance as large as $\rm 10\%$ of either 0.1 or $\rm 10\mu m$ iron particles to $\rm 10\mu m$ olivine (models 23 and 24) has \changes{little effect} on the rim location (model 1), as the cooling efficiency of larger olivine grains is similar to that of iron.

	Model 25 shows the effects of a mass abundance of 0.0001 of $\rm 0.1\mu m$ corundum grains in a disc of olivine of the same size. \changes{The corundum is significantly cooler than the olivine, as seen in Fig.~\ref{fig:3maps-oc}. The rim has moved from 2.22AU (the $\rm0.1\mu m$ olivine rim in model 0) to 1.57AU, with an optically thin zone beginning already at 0.84AU, i.e. covering $\rm\approx 45\%$ of the radial distance and $\rm 70\%$ of the surface area inside the $\rm\tau=1$ radius.}

	Allocating corundum a mass fraction of 0.01 (model 26), a rough upper limit based on the assumption that the cosmic abundance of aluminium is all in corundum, \changes{demonstrates that this species can draw the rim in considerably. Small, $\rm 0.1\mu m$ corundum grains draw the rim in more efficiently than iron of the same size, both because they are relatively more refractory at the low densities they have in these models (as low as $\rm 10^{-17}g/cm^{3}$, see Fig.~\ref{fig:sublimation}), and because they cool more efficiently at temperatures typical of dust sublimation}.

	\changes{A comparison of how small abundances of $\rm0.1\mu m$ corundum and iron in a disc of identically sized olivine cool the inner disc, and how their refractivity allows them to exist closer to the star, is presented as temperature maps of models 25, 21 and 22 in Fig.~\ref{fig:3maps-oc}. Corundum (top row) cools more efficiently than iron (e.g. middle row), and with a further contribution from its high refractivity, it moves the rim to 1.57AU, closer to the star than an identical mass abundance of iron, which is at 1.63AU. A $\rm 10\%$ mass abundance of small iron grains is sufficient for them to dominate the rim structure and decrease the radius considerably, to 0.71AU. Due to the similar cooling properties of $\rm 0.1\mu m$ iron and $\rm 10\mu m$ olivine grains, models 15 and 22 are similar.}

\section{Discussion}\label{sec:discussion}

	\subsection{Modelling summary}

	We have successfully employed the Monte Carlo radiative transfer code MCMax \citep{Minetal2009} in modelling the inner rim structures of dusty passive protoplanetary discs.

	Results in Section~\ref{sec:results} show that, especially for larger grains, a rigorous treatment of dust sublimation and condensation, together with other rim physics in MCMax, may not always lead to a stable solution for a solid dust density and temperature distribution in a passive, static disc framework.

	\subsection{Where is the rim?}\label{sec:whereistherim}

	The rims of MCMax disc models are generally between the optically thin destruction and the backwarmed wall radii (the $\rm C_{bw}=1$ and 4 radii, respectively). Grains can begin to attenuate the stellar radiation field from the $\rm C_{bw}=1$ location outward, allowing the rim to exist closer than a fully backwarmed estimate. That the rim location could vary like this has been pointed out earlier \citep[e.g.][]{Monnieretal2005}, but we have explored the phenomenon in detail for the first time.

	For a given dust species, larger grains will generally cool more efficiently and thus will set the rim location, but only if they are present in sufficient abundance, as shown in Figures~\ref{fig:twospecies}~and~\ref{fig:3maps-oo}. Large grains of olivine can exist much closer to the star than small grains. A small, 0.0001 abundance of $\rm 10\mu m$ grains is enough to create an extended, cool optically thin region in front of the rim, which is still determined by $\rm 0.1\mu m$ grains. Allocating a fraction 0.1 of the mass to large grains is sufficient for them to determine the rim location.

	\changes{It was noted by IN05 that the most refractory species would determine the rim location, if it was also able to build up enough optical depth. They also pointed out that for one species, the largest grains would position the rim under the same assumption. We add the generalization that refractivity and cooling efficiency combine to determine the species which sets the rim radius. A very high cooling efficiency may allow a species which does not have the highest sublimation temperature to determine the rim location.}

	For a given cooling efficiency, the species with the highest sublimation temperature will determine the rim location, again assuming a sufficient abundance of it is present. Corundum has been considered an ultra-refractory condensate in discs, which it is if one assumes that all species have equal partial pressures. However, if the dust vapour densities are taken equal, as in vertical slices in Fig.~\ref{fig:sublimation}, corundum is more refractory at vapour densities $\rm\rho_{vapour}<10^{-10}$, but at $\rm\rho_{vapour}>10^{-10}$, the sublimation temperature of iron is higher and increases rapidly. Hence, iron is the most likely dust species to be responsible for rim temperatures above 1500K
	\footnote{This is subject to the validity of the thermodynamic data for iron from P94. Sublimation temperatures more than 200K lower than this are found with data from \citet{CameronFegley1982}. However, P94 considered a protoplanetary nebula environment, where iron-rich silicates and other species contribute to the vapour pressure of iron. They assumed a silicate to iron mass ratio similar to ours, therefore we use their results with confidence but state that a full treatment of the multi-species gas-grain equilibria is desired.}.

\begin{figure}
		\includegraphics[bb=2.5cm 1.75cm 19cm 25cm, angle=90, clip=, width=1.0\linewidth]{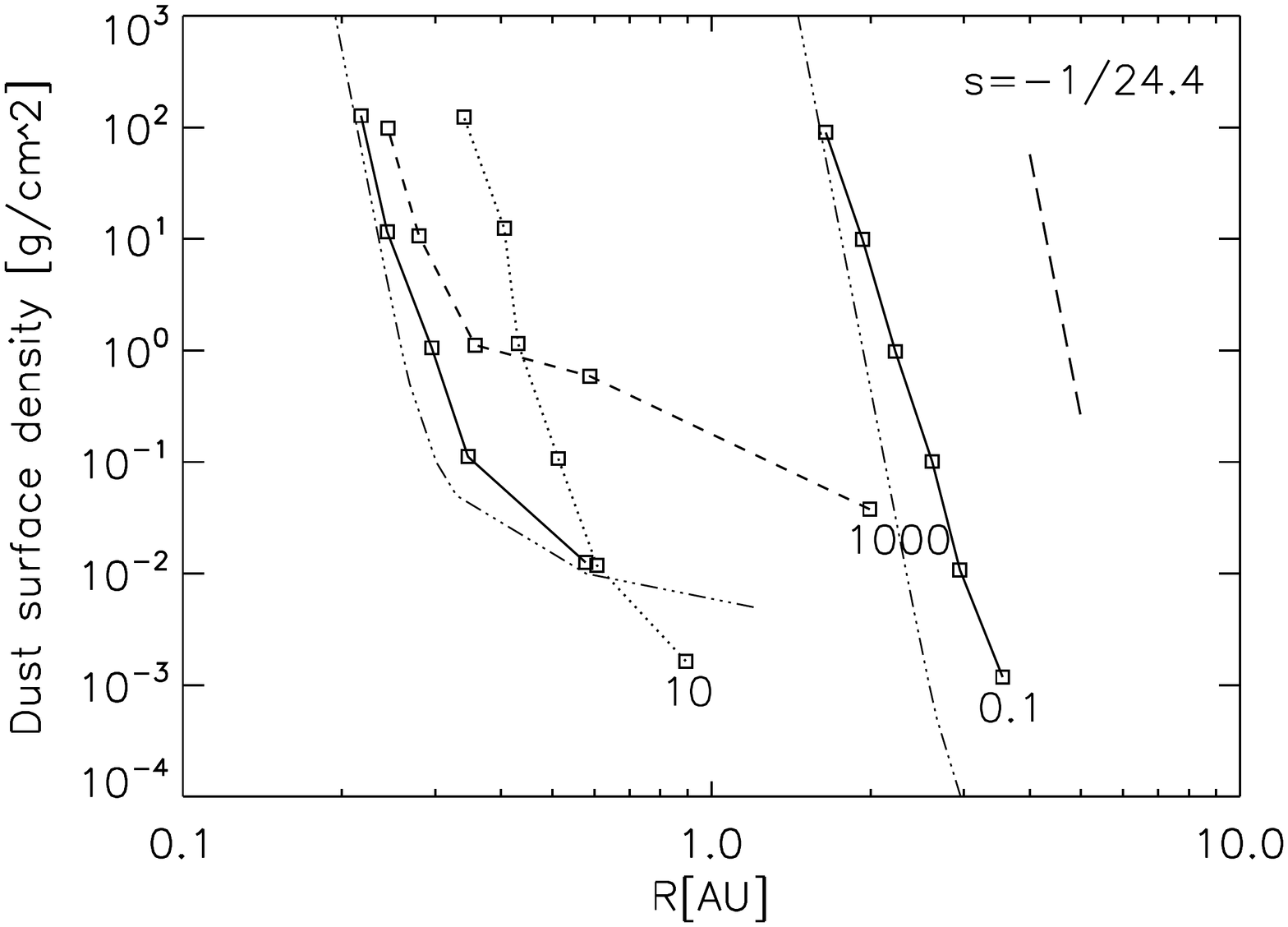}
		\caption{\changes{Midplane rim locations and dust surface densities for various sizes of olivine. \textit{Boxes} represent the final computed MCMax iteration, usually between 50 and 200. The left \textit{solid} line connects models with $\rm100\mu m$ grains, and the right $\rm0.1\mu m$. \textit{Dash-dotted} lines represent analytical computations for the same grain sizes with Eq.~\ref{eq:rimR} for $\rm C_{bw}=1$ and a dust temperature of 1500K. The \textit{dotted} line connects models with $\rm10\mu m$ grains, and the \textit{dashed} one $\rm 1000\mu m$. \textbf{Grain size variations:} The rim radius $\rm R_{rim}$ changes by almost an order of magnitude for olivine grain size variations. \textbf{Surface density variations:} For a given grain size, increasing the surface density decreases $\rm R_{rim}$, in one limit as a power law (shown by the \textit{long-dashed} line). Models for a given grain size follow irregular lines due to variations between iterations (see Fig.~\ref{fig:iterations}).}}
	\label{fig:temp-oA}
	\end{figure}

	The above points to the fact that if grain growth proceeds steadily and the dust has a complex composition featuring efficiently cooling and refractory metal oxides such as iron or corundum, the rim will exist close to the star, at 0.4AU or closer for a $\rm 47L_{\odot}$ Herbig star if $\rm\Sigma_{d} \geq 10^{0} g/cm^{2}$ (see Figures~\ref{fig:twospecies}~and~\ref{fig:temp-oA}). Evidence for grain growth is presented in e.g. \citet{vanBoekeletal2004} and \citet{Herbstetal2008}. Relatively large rim radii, for a star of the type used here $\rm R_{rim}>2AU$, are likely to arise as the inner disc is cleared of gas by the magneto-rotational instability \citep{Chiangetal2007} or a planet, lowering the sublimation temperature. Dust may also be dynamically cleared.

	\changes{Surface density controls the rim location for a given grain size. This is described by a $\rm R_{rim}\propto \Sigma_{dust}^{s=-1/24.4}$ power law in the sublimation-controlled regime and is illustrated in Fig.~\ref{fig:temp-oA}. The low density ends of lines in Fig.~\ref{fig:temp-oA} branch off from the power law as there is insufficient mass to reach $\tau = 1$ at the dust evaporation location. An analytical description of the rim location as a function of surface density for both the sublimation-controlled and optical depth controlled regimes is given in Eq.~\ref{eq:rimR}. It is a curve which branches from the power law and asymptotically becomes parallel to the x-axis.}

	\begin{equation}
	\begin{aligned}
		\rm R_{rim} = \left[ C_{1}^{-p} \left(22.09\Sigma_{subl}^{0.020}S_{1} \left(\frac{C_{bw}}{\epsilon}\right)^{-0.252}\right.\right.-\\
			\rm \left.\left. C_{2}\frac{1+4p}{\kappa^{\prime}}\Sigma_{subl}^{-1-0.078p}S_{2} \left(\frac{C_{bw}}{\epsilon}\right)^{\frac{1+8.080p}{8}}\right)\right]^{-\frac{4}{1+4p}}[cm]
	\label{eq:rimR}
	\end{aligned}
	\end{equation}

	This equation, derived in Appendix~\ref{sec:analytical}, yields the curves seen in Fig.~\ref{fig:temp-oA}. $\rm\kappa^{\prime}$ is the dust opacity at $\rm 0.55\mu m$, in $\rm cm^{2}/g$. Note that the surface density profile is anchored to $\rm\Sigma_{subl}$, its value at the sublimation radius, with a radial dependence of $\rm\Sigma\propto R^{-p}$. The properties of the central star are hidden in $\rm S_{1}~=~M_{\star}^{0.010}~L_{\star}^{-0.252}$ and $\rm S_{2}~=~M_{\star}^{-\frac{1+0.078p}{2}}~L_{\star}^{\frac{1+8.080p}{8}}$. The constants are given by $\rm C_{1}~=~4.1982\cdot 10^{-6}$ and $\rm C_{2}~=~4.84\cdot10^{8}$.

	\changes{The good correspondence of the analytical curves described by the above equation with our numerical results shows that Eq.~\ref{eq:rimR} captures the processes that are thought to dominate the rim location in dusty discs, and we propose it as an aid in interpreting more sophisticated numerical models, but also observations.}

	The sublimation properties of olivine were used in the above derivation. To use the presented formalism for any dust species, a power law fit to \cite{P94} data should be obtained ($\rm T_{subl}~=~G_{K}~(\rho_{gas}~/~[1g/cm^{3}])^{\gamma}$, see also IN05 and compare with Eq.~\ref{eq:CCevaplaw}).

	Omittances in the above analytical approach include the temperature dependence of the cooling efficiency, $\rm\epsilon(T_{dust})$, and the fact that partial condensation of dust occurs in front of the $\rm\tau=1$ location, i.e. the rim is always somewhat diffuse. These two effects are the main culprits in the discrepancies between the numerical and analytical curves seen in Fig.~\ref{fig:temp-oA}.

	We use a flat dust surface density profile $\rm\Sigma\propto R^{-p}$, where $\rm p=0$, in the inner disc, however Eq.~\ref{eq:rimR} is generic and allows to use the canonical $\rm p=1\ldots 1.5$. Furthermore, the sublimation location for a given surface density and grain type is independent of the surface density power law.

	\changes{Used with observed rim radii, Eq.~\ref{eq:rimR} may prove useful in estimating the surface density in the inner disc. For this, the equation should be solved numerically for $\rm\Sigma_{subl}$, inserting the stellar properties and adopting constraints on the grain opacity and the surface density power law.}

	\subsection{Is the rim diffuse or sharp?}

	The optically thin inner disc region is an important concept. A $\rm\tau < 1$ region between $\rm C_{bw}=1$ and 4 is supported by any of three conditions: 1) a low surface density, 2) a high surface density but very low opacity, and 3) early stages of dust condensation in a cool region.

	Condition 1 can be met by a low-abundance of a species which is more refractory than the bulk of the dust, such as corundum (see model 25 in Fig.~\ref{fig:twospecies}), or which can cool relatively efficiently, such as large grains or iron (models 14 and 21 in Fig.~\ref{fig:twospecies}).

	Condition 2 could be met by a dust type which is very transparent in the optical, but cools efficiently in the NIR. If the scattering phase function of such a species is not strongly forward-peaked, a population of grains closer to the star than the rim (determined in this case by the sublimation of another, more opaque species) could further give rise to an extended hot radiation zone through scattering.

	Condition 3 is met as a transient phenomenon in our models, which allow dust to slowly condense into an initially empty inner disc, creating and then filling in an optically thin region, as illustrated by Fig.~\ref{fig:iterations}. Transient heating events such as a powerful flares may destroy dust in an inner disc region, which will subsequently go through a similar cycle of a transient extended optically thin zone during dust re-condensation. It would be surprising to observe long-lived sharp rims composed of only large grains. However, our modelling also indicates that adding several types of dust to a model makes the rim structure much more stable.

	\subsection{Dust differentiation}

	Differentiation of dust types can occur in the optically thin region, where highly refractory or very efficiently cooling species can exist independently of more volatile or less efficiently cooling species. Maintaining such an extensive optically thin region in a disc with a diverse and broad distribution of grain sizes and compositions requires considerable fine-tuning, thus sublimation-based differentiation in a static disc is unlikely.

	\subsection{Observational implications}

	\changes{The smallest observed characteristic rim radii for $\rm L_{\star}\approx 50L_{\odot}$ stars are around 0.2AU \citep{MillanGabetetal2007}. These come from ring model fits to visibility curves. In the framework of our models, such small radii require grains of around $\rm 100\mu m$ to be present at dust surface densities of $\rm\Sigma_{d}\approx 10^{0\ldots2}g/cm^{2}$ (Fig.~\ref{fig:twospecies}, in particular models 11 to 13).}

	\changes{It may be feasible to put a lower limit on the inner rim surface density $\rm\Sigma_{rim}$ of a system with Eq.~\ref{eq:rimR}, under the assumptions that a measured $\rm R_{rim}$ corresponds to $\rm C_{bw}\approx 1$, and that $\rm R_{rim}$ is determined by the sublimation of grains with $\rm\epsilon\approx 1$. The former is suggested by comparison of the Monte Carlo and analytical results in Fig.~\ref{fig:temp-oA}, and the latter reflects the assumption that large, efficiently cooling grains will dominate the inner disc. One could then use a measured $\rm R_{rim}$ with the stellar luminosity and mass to obtain a simple lower limit on the surface density.}

	\changes{Furthermore, it is interesting to assume that the largest grains needed to satisfactorily fit the mid- to far-infrared SED of a system provide a lower limit on the size of the grains setting the location of the inner rim. Under this assumption, an upper limit for $\rm\Sigma_{rim}$ could be obtained by using the cooling efficiency $\rm\epsilon$ of these grains in Eq.~\ref{eq:rimR}. If we further assume that these grains dominate the rim location instead of $\rm\epsilon\approx1$ grains, we are again left with a lower limit on $\rm\Sigma_{rim}$.}
	
	Comparing these limits with observed disc masses and canonical $\rm\Sigma\propto R^{-1\ldots-1.5}$ power laws will constrain the difference between the inner and outer disc surface density profiles.

	Our results hold if the optical depth of gas in the inner hole to stellar radiation is $\rm\tau_{g,\star} \ll 1$ and, relatedly, if accretion can be neglected. If the gas continuum opacity in the inner hole is $\rm\tau_{gas,\star}\approx 1$, shielding will decrease the dust destruction radius by $\rm[\exp{(-1)}]^{1/2}\approx 20\%$, if the gas itself does not approach stellar temperatures. A hot gaseous component may give rise to NIR emission observed inside the dust destruction radius in some systems \citep{Eisneretal2007, Tannirkulametal2008a, Tannirkulametal2008b, Krausetal2008, Isellaetal2008}.

	\changes{As \citet{Najitaetal2009} have pointed out, modelling indicates that gas in the inner hole of a Herbig Ae/Be star should produce spectral features of CO and H$_{2}$O, but instead a continuum process was found necessary to explain the compact NIR excess of MWC480. Potentially relevant opacity sources inside the classical dust sublimation radius include H$_{2}$O, $\rm H^{-}$, free-free emission and highly refractory or transparent dust species \citep[e.g.][]{Najitaetal2009}. However, even the most refractory species hypothesized to exist in discs, such as corundum, are not expected to have a significant solid fraction inside the classical silicate dust sublimation radius.}

	\changes{If a dust species is responsible for a extended hot emission component reaching from a few stellar radii to the optically thick dust rim, it is likely to have low absorption in the visible range and a high cooling efficiency in the NIR. Moreover, species with small absorption cross-sections and relatively isotropic scattering phase functions in the visible or NIR regimes may play a role in creating an extended radiating zone in the inner disc by scattering stellar or rim emission.}

	\changes{Forsterite meets some of the above requirements, however it is not able to cool efficiently in the NIR and thus differs little from olivine in terms of rim structure, as seen in Fig.~\ref{fig:twospecies}.}

	\subsection{Looking ahead}

	\changes{To make full use of our work, observed SEDs and interferometric visibilities (as opposed to rim radii derived from them under various assumptions) need to be fitted simultaneously with MCMax or a similarly capable code.	A paper focussing on the observational applications of our models is in preparation. We are performing a study of the extreme values of observables that can be obtained with static dust models, as well as simultaneously fitting the SEDs and visibilities of specific objects.}

	We are actively exploring whether the fractional NIR excesses of our models provide a source of additional information about the inner disc surface density and/or grain type. As $\rm f_{NIR}$ varies with iteration as well as grain, disc and stellar properties, a better handle on the variability of the rim is needed before this fraction is useful.

	Rim temperatures obtained from NIR SED-fitting and limits on inner disc gas densities would increase the usefulness of Eq.~\ref{eq:rimR} in constraining the amount and type of dust in the rim region.

	We support the conclusion, overlapping in part with \citet{Monnieretal2005}, that detailed modelling of multi-species condensation and sublimation processes, as well as simultaneous treatment of radiative transfer in the gas and dust components, is needed to make further progress on the numerical side, and encourage gas and dust modellers to join efforts.

\section{Conclusions}\label{sec:conclusions}

	We have demonstrated a wide range of possible inner rim structures in a parameter study of grain size, composition and dust surface density, as well as multiple grain type models. To do this, we implemented dust sublimation and condensation physics into a fast Monte Carlo radiative transfer code. Here, we outline our main conclusions:

	\begin{enumerate}
	\item{The inner rim in dusty discs is not an infinitely sharp wall. \changes{Backwarming effects combine with sublimation and condensation in leading to a diffuse region that can extend over a significant fraction of the rim radius. In our treatment, dust generally begins to condense near the $C_{bw}=1$ location and the $\rm \tau=1$ surface is closer to the star than in previous models.}}

	\item{The inner disc surface density \changes{of a given species is an important parameter, because it determines the highest possible temperature where the species} can be stable. High surface densities move the rim closer to the star.}

	\item{\changes{The dust component (species or size) which determines the rim location by building up an optical depth of unity closest to the star need not have the highest sublimation temperature, it may instead cool very efficiently.}}

	\item{If large particles are abundant enough to produce optical depth, then the rim location will be dominated by silicates. If large particles are not abundant enough, corundum and iron grains will likely set the rim location. In either case, for our standard star, the rim location is typically around 0.4 AU, much smaller than models for $\rm 0.1\mu m$ silicate grains would predict.}

	\item{For any given grain material and size, the rim location can be analytically expressed as a function of stellar properties and of surface density. \changes{We give this expression in full form for a silicate-dominated rim.}}

	\item{The optically thin region can cover \changes{$\rm 70\%$} of the surface inside the rim radius in cases where a very efficiently cooling or highly refractory species is present in low abundance compared to the less efficiently cooling or more volatile bulk of the dust.}

	\item{In the case of very low surface densities, the build-up of optical depth will be slow, and the rim region will be very diffuse.}
	\end{enumerate}

\appendix

\section{Analytical relation between rim radius and surface density}\label{sec:analytical}

	The rim location for a given grain type is a function of surface density at the rim location, as seen in Fig.~\ref{fig:temp-oA}. At high surface densities, this dependence is a power law. This power law relation is a limiting case of a more general relation describing the $\rm\tau=1$ location as a function of surface density. At low surface densities, the rim location moves away from the dust sublimation radius as not enough dust is available there to reach an optical depth of unity.

	The calculations that follow use the \texttt{cgs} system of units. This is important to keep in mind when using the final expressions, where everything is implicit except the relevant characteristics of the dust, the disc and the star.

\subsection{Sublimation-controlled regime}

	We assume initially that the rim is at the dust destruction location, $\rm R_{rim}\equiv R_{subl}$, and wish to describe this radius as a function of dust surface density, $\rm\Sigma_{d} = \Sigma_{gas}/f_{gd}$, where $\rm f_{gd}=100$ is the gas to dust ratio. Assuming the minimal rim radius is at the dust sublimation location, with $\rm C_{bw}=1$, we first express the temperature of a grain at the rim radius from Eq.~\ref{eq:BW}:

	\begin{equation}
		\rm T_{d} = \left(\frac{C_{bw}}{\epsilon}\right)^{\frac{1}{4}} \left(\frac{L_{\star}}{16\pi\sigma}\right)^{\frac{1}{4}}R_{subl}^{-\frac{1}{2}},
	\label{eq:ATd}
	\end{equation}

	where $\rm C_{bw}$ is the backwarming factor, $\rm\epsilon$ the cooling efficiency of the grain and $\rm L_{\star}$ the luminosity of the star. This expression assumes the disc is optically thin until the rim radius.

	Further assuming that the gas and dust are in thermal equilibrium, their scaleheights will be the same and we can relate the midplane dust density $\rm\rho_{d}$ to the dust surface density via

	\begin{equation}
		\rm\Sigma_{d}(R_{rim}) = \int_{-\infty}^{\infty}\rho_{d}(R_{rim})\exp{\left(-\frac{z^{2}}{2h^{2}}\right)}dz,
	\label{eq:Asdens}
	\end{equation}

	where z is the vertical distance from the midplane and h is the scaleheight, $\rm h\propto T_{d}^{1/2}R_{rim}^{3/2}$. Integrating Eq.~\ref{eq:Asdens} and replacing the temperature with Eq.~\ref{eq:ATd}, we obtain a relation connecting $\rm\Sigma_{d}(R_{rim})$, $\rm\rho_{d}(R_{rim})$ and $\rm R_{rim}$.

	Using the IN05 power law fit to the sublimation law of olivine, we can write

	\begin{equation}
		\rm T_{d}\equiv T_{subl}=2000\left(\frac{\rho_{d}f_{gd}}{a_{d}\cdot1g/cm^{3}}. \right)^{0.0195}
	\label{eq:AsubT}
	\end{equation}

	Here, the bulk gas density, used by IN05 in fitting, has been broken into the dust density $\rm\rho_{d}$, abundance of the species $\rm a_{d}$ and the gas to dust ratio $\rm f_{gd}$. Taking the appropriate abundance from P94, we obtain the relation:

	\begin{equation}
		\rm R_{subl} = C_{1} \frac{L_{\star}^{0.53}}{\Sigma_{d}^{0.04}M_{\star}^{0.02}}\left(\frac{C_{bw}}{\epsilon} \right)^{0.53} [cm]
	\label{eq:Aplaw}
	\end{equation}

	This power law, where $\rm C_{1}~=~4.198\cdot 10^{-6}$, describes the minimum radius where dust can condense as a function of the surface density at that location. Note that if one assumes full condensation of the dust, backwarming will move the rim further, an effect which can be explored by varying $\rm C_{bw}$. A more general expression for the rim location is formulated below.

\subsection{Optical depth controlled regime}

	The inner rim is defined as the radial $\rm\tau=1$ location for a stellar photon at $\rm\lambda=0.55\mu m$. For high surface densities and under the assumptions of full condensation and $\rm C_{bw}=1$, this corresponds closely to the dust destruction radius. However, the lower the surface density, the longer the distance along the midplane that photons have to travel to reach an optical depth of unity. Thus, at relatively low surface densities, one predicts (and our numerical results show) a turn-off from the power law of Eq.~\ref{eq:Aplaw}.

	We begin by integrating the optical depth through dust radially along the midplane:

	\begin{equation}
		\rm \tau = \kappa^{\prime}\int_{R_{subl}}^{R_{rim}}\rho_{d}(R)dR.
	\label{eq:Atau}
	\end{equation}

	Here, $\rm R_{subl}$ is the dust destruction radius of Eq.~\ref{eq:Aplaw}, $\rm R_{rim}$ is the rim radius, for which one assumes $\rm\tau=1$, and $\rm\kappa^{\prime}$ is the mass absorption coefficient of the dust at $\rm 0.55\mu m$. $\rm\rho_{d}(R)$, the radial run of dust density, is obtained by multiplying Eq.~\ref{eq:Asdens} with a factor $\rm\Sigma_{subl}(R_{subl}/R)^{-p}$. This is a surface density power law fixed at the sublimation location, i.e. $\rm\Sigma_{subl}$ and $\rm R_{subl}$ obey Eq.~\ref{eq:Aplaw}. Assuming some power p relates the rim radius to the surface density at the sublimation location.

	Employing Eq.~\ref{eq:Aplaw} for $\rm R_{subl}$ and inserting physical constants, we are left with the following description of the rim location:

	\begin{equation}
	\begin{aligned}
		\rm R_{rim} = \left[ C_{1}^{-p} \left(22.09\Sigma^{0.020}S_{1} \left(\frac{C_{bw}}{\epsilon}\right)^{-0.252}\right.\right.-\\
			\rm \left.\left. C_{2}\frac{1+4p}{\kappa^{\prime}}\Sigma^{-1-0.078p}S_{2} \left(\frac{C_{bw}}{\epsilon}\right)^{\frac{1+8.080p}{8}}\right)\right]^{-\frac{4}{1+4p}}[cm]
	\label{eq:ArimR}
	\end{aligned}
	\end{equation}

	This equation yields the curves seen in Fig.~\ref{fig:temp-oA}. P relates to the surface density power law as $\rm P=(1+4p)/4$ and the properties of the central star are hidden in $\rm S_{1}~=~M_{\star}^{0.010}~L_{\star}^{-0.252}$ and $\rm S_{2}~=~M_{\star}^{-\frac{1+0.078p}{2}}~L_{\star}^{\frac{1+8.080p}{8}}$. The constants are given by $\rm C_{1}~=~4.198\cdot 10^{-6}$ and $\rm C_{2}~=~4.84\cdot10^{8}$. It is important to keep in mind that the sublimation properties of olivine were used in the above derivation.

\section{Differences from IN05 and \citet{Tannirkulametal2007}}\label{apx:differences}

	\changes{In modelling discs parametrized identically with those of \citet[IN05]{IN05}, we find systematically smaller rim radii. The main reasons are that the IN05 formalism yields a higher optically thin dust temperature and a lower sublimation temperature than the treatment used in this work, as discussed below. A similar study by \citet[][T07]{Tannirkulametal2007}, which we also comment on, found results different from both this work and IN05.}

\begin{figure}
		\includegraphics[bb=2.5cm 1.75cm 19cm 25cm, angle=90, clip=, width=1.0\linewidth]{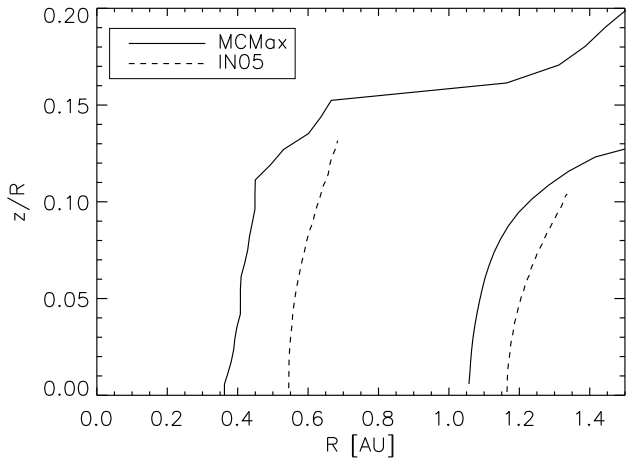}
	\caption{\changes{Radial $\rm\tau=1$ results for the IN05 disc models (\textit{dashed lines}) and the same discs modelled with MCMax (\textit{solid lines}). Grain size goes from left to right as 1.3 and $\rm 0.1\mu m$. A systematic difference in radius is evident.}}
	\label{fig:in05rims}
	\end{figure}

	\changes{We find rim radii smaller than those of both IN05 and T07 and for all modelled grains, i.e. 0.1, 0.5 and $\rm 1.3\mu m$ astronomical silicate. As seen in Fig.~\ref{fig:in05rims}, the relative difference increases with grain size, from $\rm\Delta R_{rim}\approx 10\%R_{IN05}$ for $\rm 0.1\mu m$ grains to $\rm \Delta R_{rim}\approx 35\%R_{IN05}$ for $\rm 1.3\mu m$.}

	\changes{The IN05 disc model uses the radiative transfer equation given by \citet{Calvetetal1991} for a semi-infinite slab to get a radial temperature structure inside the rim, and then calculates the vertical structure as in CG97 and DDN01. A gas to dust ratio of 100 is used, and the total disc mass is distributed according to a $\rm \Sigma\propto R^{-1.5}$ surface density power law extending from $\rm R_{in}=0.1AU$ to $\rm R_{out}=200AU$. The dust opacity determines the temperature and vertical structures, and the gas density is used to compute the dust sublimation temperature and thus governs dust sublimation.}

	\changes{To obtain a law for the dependence of the dust sublimation temperature on gas density, IN05 fitted Eq.~\ref{eq:AsubT} to data from P94. The fitted points were $\rm T_{subl}$ of olivine (olivine is used interchangeably with astronomical silicate in this appendix, as the two are assumed to differ in optical, and not in sublimation properties) and $\rm\rho_{gas}$, the bulk gas density. Because the grain-gas equilibrium is maintained by the partial pressure of olivine, this approach yields a $\rm T_{subl}=T(\rho_{gas})$ law which is applicable only to gas of the same composition as that of P94. By applying this law to a disc with a purely olivine dust composition, IN05 obtained lower sublimation temperatures than us, as discussed next.}

	\changes{P94 supplied the fractional mass abundances of the species in their gas. For olivine, it is 0.00264, i.e. $\rm 26.4\%$ of the dust mass, assuming a gas to dust ratio of 100. Thus, in the IN05 approach, olivine contributes $\rm 100\%$ of the dust mass, but the sublimation temperature is computed from the bulk gas density under the assumption that there is $100/26.4$ times less olivine present, leading to a correspondingly lower $\rm T_{subl}$. This underlies our choice to reduce the sublimation law to a dependence of $\rm T_{subl}$ on the vapour density of the species, $\rm\rho_{vapour}$, and not the bulk gas -- by doing so, we may use the law for all nebular compositions\footnote{This is valid under the assumption that olivine is the only species contributing to its own gas pressure. While this is not the case as there is no gas phase olivine molecule, it is the best we can do at present if we wish to model discs with a composition different from the P94 nebula.}.}

	\changes{Fitting the P94 data for the olivine gas density removed its abundance from our $\rm T_{subl}$ law, allowing us to compute the olivine sublimation temperature for any mass abundance. In an identical disc with pure olivine dust and $\rm f_{gd}=100$, we obtain higher sublimation temperatures than IN05, because for the same bulk gas density, pure olivine can maintain equilibrium at a higher temperature than that given by the law IN05 used.}

\begin{figure}
		\includegraphics[bb=2.5cm 1.75cm 19cm 25cm, angle=90, clip=, width=1.0\linewidth]{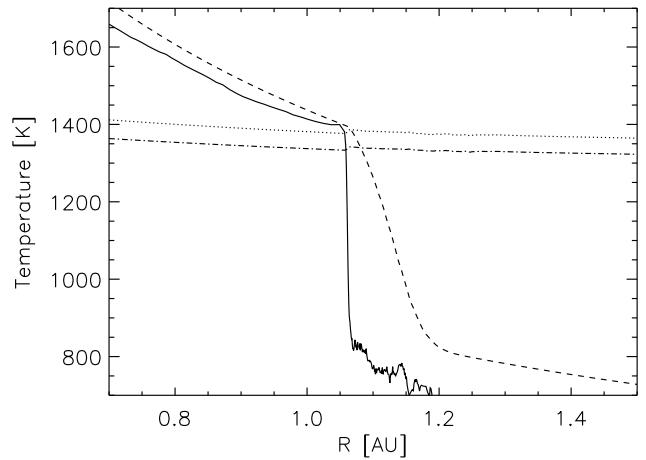}
	\caption{\changes{Midplane dust temperature structures for an MCMax model of the IN05 disc with $\rm 0.1\mu m$ astronomical silicate grains. The MCMax dust temperature structure (solid line) is seen to drop more rapidly, and is significantly cooler, in the rim region than the estimate obtained for the same density structure with the radiative transfer equation used by IN05 (dashed line). The $\rm\sim 50K$ difference between our estimate of the dust sublimation temperature (dotted line) and that of IN05 (dash-dotted line) is also clear. In the MCMax model, the rim radius is at 1.05AU. The intersection of the extrapolated optically thin temperature curves and the $\rm T_{subl}$ computed by IN05 is at $\rm\sim1.16AU$. Inside the rim, the MCMax dust temperature and that computed using the IN05 radiative transfer equation coincide better in later iterations, when more dust has condensed there and made the rim structure more similar to a wall.}}
	\label{fig:in05Tcomp}
	\end{figure}

	\changes{The above is illustrated by Fig.~\ref{fig:in05Tcomp}, which gives various temperatures in the rim region of an IN05 comparison disc with $\rm 0.1\mu m$ grains, modelled with MCMax. The dash-dotted line is the sublimation temperature computed and used in MCMax, according to our fit to P94 data. The dotted line, seen to lie $\rm\sim 50K$ below the previous, shows $\rm T_{subl}$ calculated from the same density structure using the IN05 sublimation law.}

	\changes{To use the IN05 law generally, a ``dummy'' bulk gas density can be calculated from an olivine density using both the P94 olivine abundance of 0.00264 and $\rm f_{gd}=100$. This density can be used in Eq.~\ref{eq:AsubT} to obtain $\rm T_{subl}$. We do this when using Eq.~\ref{eq:AsubT} in Appendix~\ref{sec:analytical}, obtaining results which match our numerical models very well.}

	\changes{For radiative transfer, IN05 adopted the approach of \citet{Calvetetal1991}. The disc is assumed to be an irradiated semi-infinite slab. A temperature structure derived from this overestimates the disc temperature, because a semi-infinite slab can only cool in one direction, but a disc has a finite height. Two dust temperature profiles are given in Fig.~\ref{fig:in05Tcomp}, one from MCMax (solid line) and the other computed for the same density structure using Eq.1 of IN05 (dashed line). The IN05 approach overestimates the temperature inside the rim compared to a Monte Carlo computation. The backwarming factor as defined in Sec.~\ref{sec:backwarming} is useful in understanding this, as we will now discuss.}

	\changes{The expression derived by IN05 (their Eq.6) from C91 for the optically thin dust temperature is}
	\begin{equation}
		\rm T^{4}(\tau_{d}=0)\equiv T_{0}^{4}=\left(2\mu +\frac{1}{\epsilon} \right)\left( \frac{R_{\star}}{2r} \right)^{2}T_{\star}^{4},
		\label{eq:IN05thinT}
	\end{equation}

	\changes{where $\rm T_{0}$ is the dust and $\rm T_{star}$ the stellar temperature, r the distance from the star, $\rm R_{\star}$ the stellar radius, $\mu$ is the sine of the incidence angle of starlight on the rim, and $\epsilon$ the cooling factor as defined before.}

	\changes{Comparing Eq.\ref{eq:IN05thinT} to Eq.\ref{eq:BW}, taking $\mu\approx 1$, we see that a backwarming factor of $\rm C_{bw}=2\epsilon + 1$ is implicit in the IN05 calculation. Because $\rm R_{rim}\propto \sqrt{C_{bw}}$, the IN05 sublimation locations can be further than our estimates by up to a factor $\rm\sqrt{2\epsilon+1}$ for any given grain type \citep[to relate this to rim radii derived from interferometry, see Eq.5 and 6 of][]{Isellaetal2006}. Using $\rm C_{bw}(\epsilon)$ in Eq.~\ref{eq:ArimR} decreases differences with IN05 considerably (from $10\%$ to $4\%$ of $\rm R_{IN05}$ for $\rm 0.1\mu m$ grains). However, as the comparison of results from such a $\rm C_{bw}$ and our Monte Carlo calculations has shown, this assumption generally overestimates the backwarming factor at the sublimation location.}

	\changes{Midplane temperatures of a disc parametrized like an IN05 model with $\rm 0.1\mu m$ astronomical silicate are presented in Fig.~\ref{fig:in05Tcomp}. Extrapolating the optically thin dust temperature curves (solid and dashed lines) to an intersection with the IN05 sublimation temperature curve for the same density structure (dotted line), and taking into account the overestimated backwarming at the sublimation location, one sees how in MCMax, the inner rim is situated at 1.04AU and how it would be close to 1.16AU under the radiative transfer and dust sublimation assumptions of IN05.}

	\changes{Monte Carlo radiative transfer was used by T07 to model discs parametrized like those of IN05, obtaining rim radii larger than the MCMax results in all cases, but smaller than IN05 for $\rm 1.2\mu m$ grains, and larger than IN05 for $\rm 0.1\mu m$. The T07 shift from MCMax seems to be due mostly to their use of the IN05 dust sublimation law, i.e. a low $\rm T_{subl}$, and a more basic treatment of dust sublimation and the associated numerical instabilities (A.Tannirkulam, personal communication; see also Appendix~\ref{app:numerics}). It is not clear whether the latter accounts for the changed order of rim radii for small and large grains between IN05 and TORUS (used by T07).}

	\changes{In summary, compared to the treatment in the present work, the IN05 formalism leads to higher optically thin temperatures because of the assumptions implicit in the radiative transfer, and lower sublimation temperatures for pure olivine or astronomical silicate dust because of an inconsistent use of the dust density. These factors explain their comparatively larger inner rim radii. The work of T07 adopted the IN05 sublimation law with its implications for the rim radii, and the treatment of sublimation and condensation in TORUS is more basic than in MCMax, leading to further variations in rim location which we are presently unable to quantify.}

\section{Numerical implementation}\label{app:numerics}

	In this appendix we discuss the numerical implementation of the sublimation physics described previously. We find that great care has to be taken in order to avoid instabilities or incorrect results. The iterative scheme described below is accompanied by a careful regridding of the spatial grid after each iteration to make sure that the optical depth through the disc is always sampled properly, allowing for accurate radiative transfer. Although the negative effects of a somewhat less optimized spatial grid are not always evident at first sight, we find that the influence on the temperature structure is significant and a proper spatial grid is of crucial importance for a proper implementation of the sublimation physics.

\subsection{Iterative method}

	A straightforward iterative implementation of dust evaporation and recondensation does not always result in a stable solution for the dust density structure. This is because of the high optical depths through the disc and the highly nonlinear radiative transfer effects resulting from it. In addition, changing the dust density in one region of the disc affects the temperature structure at other places, making the problem highly non-local. For example, the condensation of a tiny fraction of the available dust material in the inner region can already make the region in which this material was condensed optically thick, enhancing the effects of backwarming and thereby lifting the temperature of the grains above the evaporation temperature again. At the same time, the region behind it is shielded and there dust can condense. This causes situations where the density flips between two configurations which are both not the equilibrium solution. Thus, care has to be taken not to condense too much material in a single iterative step. However, the other extreme, condensing only so much material such that the change in optical depth is smaller than unity, requires a massive number of iterations which is unfeasible given the computation time required per iteration.

	Therefore, one must take care that the amount of material condensed or evaporated in a single iteration is not too small or too large. We have considered several different schemes for deciding how much dust to add or remove. The scheme we found to be the best tradeoff between stability and speed decides how much dust to condense or evaporate according to how close the temperature of the dust species is to its evaporation temperature. In this way, it is possible to take large steps when the solution is far from equilibrium, while the size of the steps is automatically decreased when the solution is locally approached. The method is outlined below.

	The parameter we want to determine at every location in the disc is the \emph{gas fraction} of each species $i$, $\gamma_i$. The gas fraction is defined as the fraction of the total available material for forming dust grains of species $i$ that is in the gas phase. Thus $\gamma_i=1$ means dust species $i$ is totally evaporated, while $\gamma_i=0$ means all the material is in the solid phase.

	In order to determine $\gamma_i$ at each location in the disc we first compute the temperature structure for the initial guess of the dust density structure (see section~\ref{app:initial} below). From this we compute using Eq.~(1) the partial gas pressure needed to counter the evaporation at each location in the disc. This gives the equilibrium gas fraction at this temperature, $\gamma_{i,0}$. We also compute the sublimation temperature at each location, $T_\mathrm{subl}$ and the temperature of the grains in the optically thin approximation, $T_\mathrm{thin}$. The new gas fraction is then given by,

	\begin{equation}
		\gamma_{i,\mathrm{new}}=f_w\gamma_{i,0}+(1-f_w)\gamma_{i,\mathrm{old}},
	\end{equation}

	where the weighting factor $f_w$ is taken as,

	\begin{equation}
	f_w=\left\{ \begin{array}{ll}  
	\displaystyle{\left|\frac{T-T_\mathrm{subl}}{T+T_\mathrm{subl}}\right|^q\,\left|\frac{T_\mathrm{thin}-T_\mathrm{subl}}{T_\mathrm{thin}+T_\mathrm{subl}}\right|^q}	&	\textrm{for } T<T_\mathrm{subl}\\
	&\\
	\displaystyle{1-\exp\left\{-400\left[\frac{T_\mathrm{subl}-T}{T_\mathrm{subl}}\right]^2\right\}}	&	\textrm{for } T>T_\mathrm{subl}\\
	\end{array}\right.
	\end{equation}

	The parameter $q$ determines how fast the condensation of dust can take place and is adjusted such that the increase in optical depth up to the $\tau=1$ surface as seen by direct stellar radiation is always less than $10$\,\% (i.e. $\Delta\tau<0.1$). The second term in the equation for $f_w$ for $T<T_\mathrm{subl}$ ensures that in the region just outside the first few optical depths the condensation of material is not too fast. The temperature in these regions is quite low and if this term is not taken into account, the shielded regions will form so much dust that it will again start heating the region in front of it, causing another instability. By adding dust slowly also in these regions, the overall stability of the method is increased significantly. Removal of dust grains when the temperature is too high is done much faster than recondensation. This ensures that during the iterative process only a tiny fraction of the dust grains has a temperature above the sublimation temperature.

	\changes{Together with the sublimation and recondensation of dust in the disc, we also determine the vertical density profile of the disc from hydrostatic equilibrium.}

\subsection{Initial guess}
\label{app:initial} 

	The success of most iterative schemes depend heavily on the initial guess. This is also true for the iterative scheme described above. Especially the estimate for the location of the start of the inner rim as a function of height above the midplane is of crucial importance for the success of the computation. We compute the initial guess for the dust density structure using analytic consideration as described below.

	First we take the temperature structure as obtained by Eq.~(3) with an initial guess for $\rm C_{bw}$ (usually unity), ignoring the thermal radiation and extinction of the stellar radiation by other dust grains. This temperature structure can then be used to compute the gas fraction at each location in the disc, and from this we setup the first estimate for the density structure. We then do a full radiative transfer run through this density structure obtaining a temperature at each location. We use this to determine $\rm C_{bw}$ at the inner rim. Again we use Eq.~(3) to determine the temperature at each location in the disc using the new value for $\rm C_{bw}$. This will result in a slightly different location of the inner rim. We iterate this procedure several times to get a proper value for $\rm C_{bw}$ and thereby a good estimate for the location of the inner rim. After this we proceed with the iterative scheme as described in the previous section.

\subsection{Restrictions on the gas fraction gradients}

	The location of the inner rim as a function of height in the disc is to a large extend determined by the local backwarming efficiency and the density of dust forming material. In the midplane of the disc the density is highest which causes the sublimation temperature to be highest and the rim radius to be small at this height. However, due to the heigh density the backwarming efficiency at the midplane is also highest, which causes the rim to retreat to larger radii. When this retreating effect of backwarming wins from the effect of a higher sublimation temperature, the rim radius at the midplane will be slightly larger than just above it causing a 'hole' in the disc. Inside this hole the dust is irradiated from almost all sides, causing the effect of backwarming to increase even further and the rim to retreat even further. This is a runaway effect which, in the end, causes the entire disc to retreat to large radii. When this has happened, in front of this rim the temperatures are low and dust can start forming again. This cycle repeats and no stable solution is found.

	The above instability might be a physical effect which can emerge in a time-dependend treatment of the problem. However, we do not pursue this in this paper. Therefore, we restrict the gas fraction gradient to avoid triggering this instability. We restrict the gas fraction to be monotonously increasing with increasing height above the midplane. In this way we avoid the formation of a hole in the disc. 

	Another problem which might arise from the iterative procedure explained above is that the destruction of dust proceeds faster in the region which is effectively backwarmed than in the optically thin region in front of it. This might cause the fraction of condensed material to decrease with increasing radius. This solution turns out to be unstable, and in addition we consider it to be an unlikely physical solution. Thus, to avoid this we restrict the gas fraction to be monotonously decreasing with increasing radius.

	Note that by restricting the gas fraction in the ways described above we force the rim to have the curved shape we see in the paper. However, other structures we have found have in all cases turned out to be unstable, thus we consider the curved rim a robust solution.

\begin{acknowledgements}
	The authors thank the anonymous referee for providing constructive comments, which significantly helped to improve the paper. M.Kama gratefully acknowledges support from the Netherlands Organisation for Scientific Research (NWO) grant number 021.002.081 and from an HSP Huygens Scholarship. M.Min acknowledges financial support from NWO through a Veni grant.
\end{acknowledgements}

\bibliographystyle{aa}
\bibliography{paper1}

\end{document}